\documentclass[twocolumn,prd,aps,nofootinbib,showpacs,superscriptaddress]{revtex4}
\usepackage{multirow}
\usepackage{graphicx}
\usepackage[dvips]{color}
\usepackage{amssymb,amsmath}

\begin{document}

\title{Bubble contributions to scalar correlators with mixed actions}

\author{Ziwen Fu }
\affiliation{
Key Laboratory for Radiation Physics and Technology of  Education Ministry;
Institute of Nuclear Science and Technology, College of Physical Science and Technology,
Sichuan University, Chengdu 610064, People's Republic of China
}

\begin{abstract}
Within mixed-action chiral perturbation theory (MA$\chi$PT),
Sasa's derivation of the bubble contribution to scalar $a_0$ meson  is extended
to those of scalar $\kappa$ and $\sigma$ mesons.
We revealed that $\kappa$ bubble has two double poles and
$\sigma$ bubble contains a quadratic-in-$t^2$ growth factor
stemming from the multiplication of two double poles
for a general mass tuning of valence quarks and sea quarks.
The corresponding preliminary analytical expressions in MA$\chi$PT
with $2+1$ chiral valence quarks
and $2+1$ staggered sea quarks will be
helpful for lattice studies of scalar mesons.
\end{abstract}


\maketitle

\section{Introduction}
The nature of the lowest scalar meson is unveiled.
Nowadays, $\sigma$ meson  was established
with a low mass~\cite{Beringer:1900zz},
and $a_0(980)$ meson is experimentally-confirmed,
while $\kappa$ meson is still disputable~\cite{Beringer:1900zz}.
Moreover, there are long-lasting debates on the traits of the lowest scalar meson:
is it a conventional $\bar qq$ or tetraquarks
$\bar q\bar qqq$~\cite{Mathur:2006bs,Prelovsek:2010kg,Prelovsek:2004jp,McNeile:2006nv,Liu:2007hma,Fu:2011zz}?

The tetraquark interpretation can easily realize the experimental mass ordering
$m_{a_0(980)}>m_{\kappa}$ by lattice QCD,
whereas the conventional $\bar qq$ states hardly explain this mass ordering~\cite{Prelovsek:2010kg,Prelovsek:2008rf,Alford:2000mm,Loan:2008sd,Wagner:2013nta,Alexandrou:2012rm}.
Moreover, it is consistent with the MIT Bag Model~\cite{Gottfried:1986hc}.
Nevertheless, it is sharply criticized
for overlooking some unresolved physical issues such as
chiral symmetry breaking and non-trivial vacuum state~\cite{Gribov:1999ui}, etc.
This question can be partially reconciled if the masses of $\bar{q}q$
states with $I=1/2, 1$ and $0$ are robustly calculated on the lattice.
We here call these states as $\kappa$, $a_0$ and $\sigma$
mesons, respectively.

The $a_0$ meson was studied in staggered fermion~\cite{Bernard:2001av,Gregory:2005yr}.
The propagators are discovered to hold the states
with masses below the likely combinations of
two physical mesons~\cite{Bernard:2001av,Gregory:2005yr},
which can be nicely interpreted by
the bubble contribution to the $a_0$ correlator (for simplicity,
we call it ``$a_0$ bubble" in this work, likewise for ``$\kappa$ bubble" and ``$\sigma$ bubble"  )
derived by Sasa in MA$\chi$PT~\cite{Prelovsek:2005rf}.

We have been studying scalar meson for some time.
With $2+1$ Asqtad-improved staggered sea quarks~\cite{Bazavov:2009bb},
we obtained $\kappa$ mass: $826 \pm 119$~MeV~\cite{Fu:2011zz},
and rectified the taste-symmetry breaking by extending analyses
of $\sigma$ and  $a_0$
mesons~\cite{Prelovsek:2005rf,Fu:2011gw,Bernard:2007qf,Fu:2011zzh,Fu:2011zzl}
to $\kappa$ meson in the staggered chiral perturbation theory (S$\chi$PT)~\cite{Fu:2011xb,Fu:2013sua}.
We realized that $\kappa$ bubble should be involved
in a fit of $\kappa$ correlator for the MILC medium coarse ($a\approx0.15$~fm)
and coarse ($a\approx0.12$~fm) lattice ensembles.
Moreover, bubble contributions to scalar correlators in S$\chi$PT offers a test of lattice artifacts
due to the fourth-root trick~\cite{Prelovsek:2005rf,Bernard:2007qf}.

Additionally, S$\chi$PT predicts that
these lattice artifacts vanish in the continuum limit,
merely remaining physical thresholds~\cite{Prelovsek:2005rf,Fu:2011gw,Bernard:2007qf,Fu:2011zzh,Fu:2011zzl,Fu:2011xb,Fu:2013sua}.
To check this prediction, we specially studied scalar mesons
at a MILC fine ($a\approx0.09$ fm) lattice ensemble~\cite{Fu:2011zzl,Fu:2013sua}.
As expected, lattice artifacts are indeed significantly suppressed~\cite{Fu:2011zzl,Fu:2013sua}.

Lattice studies with staggered fermions are cheaper
than those of other fermion discretizations,
which allow lattice studies
with the smaller dynamical quark masses or finer lattice spacings.
But, this benefits accompanies by an extra theoretical complications.
Each staggered quark exists in four tastes~\cite{Susskind:1976jm},
and staggered meson comes in sixteen tastes,
and the taste symmetry breaking at $a\neq 0$
gives rise to discretization errors of ${\cal{O}}(a^2)$~\cite{Lee:1999zxa}.
Since these errors are usually not negligible,
these unphysical effects predicted by S$\chi$PT~\cite{Lee:1999zxa}
should  be neatly removed from the lattice data
in order to extract the desired physical quantities.

Theoretically, lattice study with domain-wall (DW) quarks
is simpler than that with staggered quarks
since they do not come in multiple species~\cite{Kaplan:1992bt,Shamir:1993zy}.
Consequently, it makes MA$\chi$PT expressions
for bubble contributions to scalar mesons simple and
continuum-like~\cite{Bar:2005tu,Prelovsek:2005rf}.
Moreover, they keep proper chiral symmetry
up to exponentially small corrections at $a\neq 0$~\cite{Antonio:2008zz}.
Nonetheless, realistic simulations with DW fermions are expensive.
We notice that there exists a pioneering study on $a_0$ meson
with  DW fermions~\cite{Aubin:2008wk}.

MA$\chi$PT for Ginsparg-Wilson type quarks on a staggered sea
was proposed in Ref.~\cite{Bar:2005tu}.
Other quantities like $a_0$ bubble have since been added in MA$\chi$PT~\cite{Tiburzi:2005is,Chen:2005ab,Chen:2006wf,Aubin:2006hg,Prelovsek:2005rf}.
It is worth mentioning that only a couple of extra parameters
enter the relevant chiral formulas.
One of these parameters was determined
by MILC Collaboration~\cite{Bernard:2001av}.
Aubin {\it et al} estimated another one inherent in the mixed-action case~\cite{Aubin:2008wk}.
These parameters can be used to study other physical
quantities: e.g., the bubble contributions to scalar mesons discussed in this work.

In this work we extend Sasa's original derivation on the bubble contribution
to $a_0$ correlator~\cite{Prelovsek:2005rf}
to those of $\kappa$ and $\sigma$ correlators
in MA$\chi$PT with $2+1$ chiral valence quarks and $2+1$ staggered sea quarks,
and exploit two above-mentioned published parameters~\cite{Bernard:2001av,Aubin:2008wk}
to elucidate our analytical expressions using two mass tuning.
We found that $\kappa$ and $\sigma$ bubbles
demonstrated many particular features,
e.g., the strong infrared-sensitive unphysical effects
due to the multiplication of two double pole to $\sigma$ bubble.
These analyses will be helpful for lattice
determinations on scalar meson masses.

This paper is organized as follows.  We review the necessary knowledge of
MA$\chi$PT in Sec.~\ref{sec:MAChPT}, and deduce $\kappa$ and $\sigma$
bubbles in MA$\chi$PT In Sec.~\ref{sec:scalarBubble}.
We illustrate the obtained analytical expressions in Sec.~\ref{sec:Num}.
The conclusions are arrived in Sec.~\ref{Sec:Conclusions}.
The corresponding S$\chi$PT $\kappa$ and $\sigma$ bubbles are
courteously dedicated to the Appendix.

\section{Mixed Action $\chi$PT}
\label{sec:MAChPT}
At leading-order quark mass expansion, the  mixed action chiral
Lagrangian is described by $N_\textrm{V}$ Ginsparg-Wilson  valence quarks
and $N_\textrm{S}$ staggered sea quarks~\cite{Bar:2005tu}.
Each staggered sea quark comes in four tastes, and each
Ginsparg-Wilson valence quark owns a bosonic ghost partner,
and these bosons are expressed by the field
$$
\Sigma = {\rm exp} ( 2 i \Phi/f ),
$$
which is an element of $U(4N_\textrm{S}+N_\textrm{V} |N_\textrm{V})$,
and $\Phi$ is a matrix gathering pseudoscalar fields.
For instance, in the case $N_\textrm{V}=3$, $N_\textrm{S}=3$~\footnote{
These fields, which are first provided to study $\pi\pi$ scattering~\cite{Chen:2005ab},
explicitly include strange valence and strange ghost quark.
},
the fields are arranged by~\cite{Bar:2005tu,Chen:2005ab}
\begin{eqnarray}\label{eq:Sigma}
\Phi \hspace{-0.1cm}&=&\hspace{-0.2cm} \left( \begin{array}{cccccccccc}
U & \pi^+ & K^+&  Q_{ux} &  Q_{uy} &  Q_{uz} & \cdots & \cdots  & \cdots\\*
\pi^- & D & K^0& Q_{dx}  & Q_{dy} & Q_{dz} & \cdots & \cdots & \cdots \\*
K^- & \bar{K}^0 & S &  Q_{sx} & Q_{sy} & Q_{sz} & \cdots & \cdots & \cdots\\*
Q_{ux}^\dagger & Q_{dx}^\dagger & Q_{sx}^\dagger &  X & P^+ & T^+
& R_{\tilde xx}^\dagger  & R_{\tilde yx}^\dagger & R_{\tilde zx}^\dagger \\*
Q_{uy}^\dagger  & Q_{dy}^\dagger  & Q_{sy}^\dagger   &  P^- & Y & T^0
& R_{\tilde xy}^\dagger  &R_{\tilde yy}^\dagger & R_{\tilde zy}^\dagger\\*
Q_{uz}^\dagger & Q_{dz}^\dagger & Q_{sz}^\dagger  & T^- &  \bar{T}^0   &
Z & R_{\tilde xz}^\dagger & R_{\tilde yz}^\dagger &R_{\tilde zz}^\dagger\\*
\cdots & \cdots  & \cdots  & R_{\tilde xx}  & R_{\tilde xy}
& R_{\tilde xz} & \tilde{X}  & \tilde{P}^+ & \tilde{T}^+ \\*
\cdots & \cdots  & \cdots  & R_{\tilde yx}  & R_{\tilde yy}
& R_{\tilde yz}  & \tilde{P}^-  & \tilde{Y} & \tilde{T}^0 \\*
\cdots & \cdots  & \cdots  & R_{\tilde zx}  & R_{\tilde zy}  & R_{\tilde zz}
& \tilde{T}^-  & \tilde{\bar{T}}^0 & \tilde{Z} \\*
\end{array}\right), \nonumber
\end{eqnarray}
where we label sea quarks by $u,d,$ and $s$
and corresponding valence quarks and valence ghosts
by $x$, $y$, $z$ and $\tilde x$, $\tilde  y$, $\tilde z$, respectively.
The sea-quark bound state fields are $U$, $\pi^{+}$, $K^{+}$, {\it etc.},
$P^+$, $T^+$, $T^0$, $X$, $Y$ and $Z$ are the $x\bar y$, $x\bar z$, $y\bar z$,
$x\bar x$, $y \bar y$ and $z \bar z$ valence bound states, respectively;
$\tilde P^+$, $\tilde T^+$, $\tilde T^0$, $\tilde X$, $\tilde Y$ and $\tilde Z$
are the relevant combinations of valence ghost quarks.
Those labeled by $R$'s are  the (fermionic) bound state composed of one valence
and one ghost quark.
Similarly, $Q_{Fv}$ stands for the bosonic mixed bound state $F\bar v$,
where $F\in\{u,d,s\}$, and  $v\in\{x,y,z\}$.
The mixed ghost-sea pseudo-Goldstone bosons indicated by ellipses
are not used in this work.

The valence-valence mesons satisfy mass relations~\cite{Bar:2005tu}:
\begin{equation}
M_{vv^\prime}^2 = \mu(m_v +m_{v^\prime}),
\label{eq:m_tree}
\end{equation}
where three valence quarks have $m_{x}$,  $m_{y}$ and $m_{z}$
(same for the corresponding valence ghost flavors), respectively.
In this work we are only interested in the degenerate up and down valence quarks
masses, to be specific, $m_x = m_y \neq m_z$ ($2+1$) case.

For a  meson of taste $b$ made up
of sea quarks $F$ and $F^\prime$ ($F\not=F^\prime$),
the tree-level results give~\cite{Lee:1999zxa}
\begin{equation}
\label{eq:SSmass}
M^2_{FF',b} = \mu (m_F + m_{F^\prime}) + a^2\Delta(\xi_b),
\end{equation}
where the staggered sea quarks $4u,4d,4s$ own masses $m_u$, $m_d$ and $m_s$, respectively,
and $\Delta_t$ is different for each of the $SO(4)$-taste
irreps:   $P$, $V$, $A$, $T$, $I$~\cite{Lee:1999zxa}.
A new operator in the mixed action Lagrangian
relates the valence and sea sectors and contributes
a taste breaking parameter $a^2\Delta_{\rm Mix}$
of the valence-sea pion mass~\cite{Bar:2005tu}.
For a $F\bar v$ meson with field $Q_{Fv}$, its mass is given by~\cite{Bar:2005tu}
\begin{equation}
M^2_{Fv} = \mu (m_F + m_v) + a^2\Delta_{\rm Mix},
\end{equation}
where parameter $\Delta_\textrm{Mix}$ can be measured via lattice QCD.

The connected propagators for valence-valence mesons
with $v,v^\prime =x,y,z,\tilde x,\tilde y,\tilde z$  are given~\cite{Bar:2005tu}
\begin{equation}
\label{mchpt_prop_vv_conn}
\langle\Phi_{vv^\prime }|\Phi_{v^\prime v}\rangle =
\frac{\epsilon_x}{k^2+M_{v,v^\prime }^2}, \quad \epsilon_{x,y,z}=1, \epsilon_{\tilde x,\tilde y,\tilde z}=-1 .
\end{equation}

MA$\chi$PT has flavor-neutral quark-disconnected
hairpin propagators involving double pole contributions.
The flavor-neutral propagators appearing in the expression for
the bubble contributions to scalar mesons
are only those with two valence quarks~\cite{Bar:2005tu},
\begin{eqnarray}
\label{mchpt_prop_vv_disc}
\hspace{-0.5cm} \langle\Phi_{vv}|\Phi_{v^\prime v^\prime }\rangle_{\rm disc}&=& \cr
&&\hspace{-0.9cm} -\frac{1}{3}\frac{(k^2+M_{U_I}^2)(k^2+M_{S_I}^2)}
{(k^2+M_{v,v}^2)(k^2+M_{v^\prime, v^\prime}^2)(k^2+ M_{\eta_I}^2)},
\end{eqnarray}
where it is convenient to use $m_0^2\to \infty$ to decouple the $\eta'_I$,
and we are only interested in $2+1$ case~\cite{Bernard:2001av},
$$
m_{\pi^0_I}^2 = m_{U_I}^2= m_{D_I}^2, \quad
m_{\eta_I}^2  = \frac{1}{3}m_{U_I}^2+ \frac{2}{3}m_{S_I}^2 ,
$$
here $M_{U_I}^2=M_{U_5}^2+a^2\Delta_I$, $M_{S_I}^2=M_{S_5}^2+a^2\Delta_I$.
It is interesting to note that the sea-sea
pseudo-Goldstone bosons in the above expressions are taste singlets.

The propagators for valence-sea mesons with $F=u,d,s$ and $v=x,y,z$
are given
\begin{equation}
\label{mchpt_prop_vs}
\langle\Phi_{v F}|\Phi_{F v}\rangle = \frac{1}{k^2+M_{v,F}^2}.
\end{equation}
It is important to note that propagators (\ref{mchpt_prop_vv_conn}),
(\ref{mchpt_prop_vv_disc}) and (\ref{mchpt_prop_vs}) rest only on
taste breaking parameters $a^2\Delta_I$ and $a^2\Delta_{\rm Mix}$.

\section{Scalar bubble term in MA$\chi$PT}
\label{sec:scalarBubble}
The simulations with chiral valence quarks on top of
the MILC staggered sea quarks is feasible and charming.
The relevant effective theory has been developed~\cite{Bar:2005tu}.
Following the original derivations and notations
in Refs.~\cite{Prelovsek:2005rf,Bar:2005tu,Chen:2005ab,Aubin:2008wk},
we here deduce the bubble contributions to the $\kappa$ and $\sigma$ correlators
in M$\chi$PT with  $2+1$ chiral valence quarks and $2+1$
MILC staggered sea quarks ($m_u = m_d \neq m_s$).
Since the relevant $a_0$ bubble contribution is derived in Ref.~\cite{Prelovsek:2005rf},
and its time Fourier transform is provided in Eq.~(11)
of Ref.~\cite{Aubin:2008wk}, in this work we will directly quote these results.

\subsection{$\kappa$ bubble}
The bubble contribution to $\kappa$ correlator is denoted in Ref.~\cite{Fu:2011xb}.
Applying the Wick contractions, we have~\cite{Prelovsek:2005rf}~\footnote{
It is interesting and important to note that the corresponding bubble contribution
to $a_0$ correlator is
\begin{eqnarray}
\label{mchpt_1}
 B^{\rm M\chi PT}_{2+1,a_0} &=& \mu^2\biggl[ \sum_{F=u,d,s}\langle\Phi_{xF}|\Phi_{Fx}\rangle\langle\Phi_{Fy}|\Phi_{yF}\rangle \cr
&&+ 2\langle\Phi_{xx}|\Phi_{yy}\rangle\langle\Phi_{xy}|\Phi_{yx}\rangle \cr
&&+ \sum_{v=x,y,z,\tilde x,\tilde y,\tilde z}
\langle\Phi_{x v}|\Phi_{v x}\rangle \langle\Phi_{v y}|\Phi_{y v}\rangle
\biggr],\nonumber
\end{eqnarray}
which results in two extra terms to original Eq.~(13) in Ref.~\cite{Prelovsek:2005rf},
which are neatly canceled each other out in the final $a_0$ bubble.
Consequently, it is nicely consistent
with Sasa's result derived with $2$ chiral valence quarks
and $2+1$ staggered sea quarks~\cite{Prelovsek:2005rf}.
}
\begin{eqnarray}
\label{mchpt_kappa}
\hspace{-0.7cm} B^{\rm M\chi PT}_{2+1,\kappa} &=& \mu^2\biggl[ 2\langle\Phi_{xx}|\Phi_{zz}\rangle\langle\Phi_{xz}|\Phi_{zx}\rangle \cr
&&+ \sum_{v=x,y,z,\tilde x,\tilde y,\tilde z}
\langle\Phi_{x v}|\Phi_{v x}\rangle
\langle\Phi_{v z}|\Phi_{z v}\rangle \cr
&&+\sum_{F=u,d,s}\langle\Phi_{xF}|\Phi_{Fx}\rangle\langle\Phi_{Fz}|\Phi_{zF}\rangle
\biggr],
\end{eqnarray}
where the third term is already considered
to reduce four tastes per sea quark to one.
The bubble contribution is secured
by inserting relevant propagators into (\ref{mchpt_kappa})
\begin{eqnarray}
\label{Bmchpt_kappa}
B_{2+1,\kappa}^{\rm M\chi PT}(p) &=& \mu^2 \sum_k\Biggl\{
-\frac{1}{(k+p)^2+M_{x,z}^2} \times \Bigg[ \cr
&&\hspace{-1.2cm}\frac{2}{3}\frac{1}{(k^2+M_{x,x}^2)(k^2+M_{z,z}^2)}
\frac{(k^2+M_{U_I}^2)(k^2+M_{S_I}^2)}{k^2+M_{\eta_I}^2}\cr
&&\hspace{-1.2cm}+\frac{1}{3}\frac{1}{(k^2+M_{x,x}^2)^2}
\frac{(k^2+M_{U_I}^2)(k^2+M_{S_I}^2)}{k^2+M_{\eta_I}^2}\cr
&&\hspace{-1.2cm}+\frac{1}{3}\frac{1}{(k^2+M_{z,z}^2)^2}
\frac{(k^2+M_{U_I}^2)(k^2+M_{S_I}^2)}{k^2+M_{\eta_I}^2}\Bigg]\cr
&&\hspace{-1.2cm}+2\frac{1}{(k+p)^2+M_{x,u}^2} \frac{1}{k^2+M_{z,u}^2} \cr
&&\hspace{-1.2cm}+\frac{1}{(k+p)^2+M_{x,s}^2} \frac{1}{k^2+M_{z,s}^2}\Biggl\} .
\end{eqnarray}
It is helpful to perform a partial fraction decomposition,
then Eq.~(\ref{Bmchpt_kappa}) can be simplified to a form
\begin{eqnarray}
\label{Bmchpt_kappa_compact}
B_{2+1,\kappa}^{\rm M\chi PT}(p) \hspace{-0.2cm}&=&\hspace{-0.2cm}
\mu^2 \sum_k\Biggl\{
-\frac{1}{(k+p)^2+M_{x,z}^2}
\times\Biggl[\frac{g_1}{k^2+M_{\eta_I}^2} \cr
\hspace{-0.2cm}&&\hspace{-0.2cm}              +\frac{g_2}{k^2+M_{x,x}^2}
              +\frac{g_3}{k^2+M_{z,z}^2}
  + \frac{g_4}{(k^2+M_{x,x}^2)^2} \cr
\hspace{-0.2cm}&&\hspace{-0.2cm}  + \frac{g_5}{(k^2+M_{z,z}^2)^2}\Biggl]
 +\frac{2}{(k+p)^2+M_{x,u}^2} \frac{1}{k^2+M_{z,u}^2} \cr
\hspace{-0.2cm}&&\hspace{-0.2cm}
+\frac{1}{(k+p)^2+M_{x,s}^2} \frac{1}{k^2+M_{z,s}^2}\Biggl\} ,
\end{eqnarray}
where
\begin{eqnarray}
g_1 &=& \frac{1}{3}\times \frac{(M_{U_I}^2-M_{\eta_I}^2)(M_{S_I}^2-M_{\eta_I}^2)}
{(M_{x,x}^2-M_{\eta_I}^2)(M_{z,z}^2-M_{\eta_I}^2)} \cr
&&\times \left[ 2 +
\frac{M_{z,z}^2-M_{\eta_I}^2}{M_{x,x}^2-M_{\eta_I}^2} +
\frac{M_{x,x}^2-M_{\eta_I}^2}{M_{z,z}^2-M_{\eta_I}^2} \right], \cr
g_2 &=& \frac{2}{3}\times\frac{(M_{U_I}^2-M_{x,x}^2)(M_{S_I}^2-M_{x,x}^2)}
{(M_{\eta_I}^2-M_{x,x}^2)(M_{z,z}^2-M_{x,x}^2)} \cr
&&+\frac{ 3M_{x,x}^2(M_{x,x}^2-2M_{\eta_I}^2)+2M_{S_I}^4+M_{U_I}^4}
{9(M_{\eta_I}^2-M_{x,x}^2)^2}, \cr
g_3 &=& \frac{2}{3}\times\frac{(M_{U_I}^2-M_{z,z}^2)(M_{S_I}^2-M_{z,z}^2)}
{(M_{\eta_I}^2-M_{z,z}^2)(M_{x,x}^2-M_{z,z}^2)} \cr
&&+\frac{ 3M_{z,z}^2(M_{z,z}^2-2M_{\eta_I}^2)+2M_{S_I}^4+M_{U_I}^4}
{9(M_{\eta_I}^2-M_{z,z}^2)^2}, \cr
g_4 &=& \frac{(M_{U_I}^2-M_{x,x}^2)(M_{S_I}^2-M_{x,x}^2)}
{3(M_{\eta_I}^2-M_{x,x}^2)}, \cr
g_5 &=& \frac{(M_{U_I}^2-M_{z,z}^2)(M_{S_I}^2-M_{z,z}^2)}
{3(M_{\eta_I}^2-M_{z,z}^2)}.
\end{eqnarray}
The time Fourier transform of this bubble contribution
(namely,
$\displaystyle B_{2+1,\kappa}^{\rm M\chi PT}(t) =
F.T.[B_{2+1,\kappa}^{\rm M\chi PT}(p)]_{\mathbf{p}=\mathbf{0}}
$)
is then provided by
\begin{eqnarray}
\label{eq:bubble_compact_kappa_t}
B_{2+1,\kappa}^{\rm M\chi PT}(t) \hspace{-0.2cm}&=& \hspace{-0.2cm}
\frac{\mu^2}{4L^3}\sum_{\mathbf{k}}
\Biggl[
-g_1\frac{e^{-(\omega_{xz} + \omega_{\eta_I}) t}}{\omega_{xz}\omega_{\eta_I}}
-g_2\frac{e^{-(\omega_{xz}+\omega_{xx})t}}{\omega_{xx}\omega_{xz}} \cr
\hspace{-0.2cm}&&\hspace{-0.2cm}-g_3\frac{e^{-(\omega_{xz}+\omega_{zz})t}}{\omega_{xz}\omega_{zz}} -g_4\frac{e^{-(\omega_{xz}+\omega_{xx})t}}{2\omega_{xz}\omega_{xx}^3}
\left(\omega_{xx} t + 1\right)  \cr
\hspace{-0.2cm}&&\hspace{-0.2cm}-g_5\frac{e^{-(\omega_{xz}+\omega_{zz})t}}{2\omega_{xz}\omega_{zz}^3}
\left(\omega_{zz} t + 1\right) \cr
\hspace{-0.2cm}&&\hspace{-0.2cm}
+  \frac{e^{-(\omega_{xs} +\omega_{zs}) t}}{\omega_{xs}\omega_{zs}}
+ 2\frac{e^{-(\omega_{xu} +\omega_{zu}) t}}{\omega_{xu}\omega_{zu}}
\Biggr],
\end{eqnarray}
where, for brevity, in this work  we use the notation
$\omega_i \equiv \sqrt{\mathbf{k}^2 + m_i^2}$ from Ref.~\cite{Aubin:2008wk}.

It is worth mentioning that no free parameters are presented in (\ref{eq:bubble_compact_kappa_t}),
which is solely predicted by MA$\chi$PT.
The meson masses and coupling constant $\mu$ are evaluated
from lattice studies~\cite{Bernard:2001av}.
The values of mixed-meson splittings $a^2\Delta_{\rm mix}$ and
taste-singlet breaking $a^2\Delta_I$
will be quoted from Refs.~\cite{Bernard:2001av,Aubin:2008wk}.
Additionally, we notice that equation~(\ref{eq:bubble_compact_kappa_t})
gets unphysical contributions from $KS$ intermediate states 
not presenting in continuum full QCD.
Luckily, it never dominates $\kappa$ bubble contribution at large $t$
even if the valence quark masses are enough small,
which make $\kappa$ mass relatively safe to be determined,
while it is difficult to extract $a_0$ mass
since unphysical $\pi\pi$ intermediate states dominate $a_0$ correlator.
Moreover, there are two double poles in its momentum-space propagator,
which lead to the infrared-sensitive linear-in-$t$ growth factors
in the fourth and fifth terms of Eq.~(\ref{eq:bubble_compact_kappa_t}).
We will observe that there is a desirable cancelation
between two double poles for a generic mass tuning.

The $KS$ intermediate states contribute to $\kappa$ bubble (\ref{eq:bubble_compact_kappa_t})
if lattice theory (e.g., MA$\chi$PT) is not unitary.
Since full QCD is restored in MA$\chi$PT only in the continuum limit,
the unphysical $KS$ contributions
can not  be entirely removed for
any selection of mixed-action realistic simulation parameters.
In  the continuum limit ($a^2\Delta_I\to 0$, $a^2\Delta_{\rm Mix}\to 0$),
the expression (\ref{eq:bubble_compact_kappa_t})
reduces to a pretty simple form
\begin{eqnarray}
\label{eq:bubble_compact_limit_kappa_t}
\hspace{-0.6cm} B_\kappa^{a=0}(t) \hspace{-0.2cm}&=&\hspace{-0.2cm} \frac{\mu^2}{4L^3}\sum_{\mathbf{k}}
\Biggl[
\frac{3}{2}\frac{e^{-(\omega_{U_5} + \omega_{K_5})t}}{\omega_{U_5}\omega_{K_5}}
\hspace{-0.015cm}+\hspace{-0.15cm}
\frac{1}{6}\frac{e^{-(\omega_{K_5} + \omega_{\eta_5}) t}}{\omega_{K_5}\omega_{\eta_5}}
\Biggr],
\end{eqnarray}
which is nicely consistent with the corresponding S$\chi$PT result,
which is written down in (\ref{BSchpt_limit_kappa}).
The desired physical contributions $e^{- (M_\pi + M_K) t}$ certainly
dominate at large $t$ with $a\neq0$.

\subsection{$\sigma$ bubble}
The bubble contribution to $\sigma$ correlator is denoted in Ref.~\cite{Bernard:2007qf}.
Applying the Wick contractions, we get~\cite{Prelovsek:2005rf}
\begin{eqnarray}
\label{Bmchpt_f0}
\hspace{-0.3cm} B^{\rm M\chi PT}_{2+1,\sigma} &=& \mu^2\biggl[
2\langle\Phi_{xx}|\Phi_{xx}\rangle\langle\Phi_{xx}|\Phi_{xx}\rangle \cr
&&
+2\langle\Phi_{xx}|\Phi_{yy}\rangle\langle\Phi_{xx}|\Phi_{yy}\rangle \cr
&&+ \sum_{v=x,y,z, \tilde x,\tilde y,\tilde z}
\langle\Phi_{x v}|\Phi_{v x}\rangle
\langle\Phi_{v x}|\Phi_{x v}\rangle \cr
&&+\sum_{F=u,d,s}\langle\Phi_{1F}|\Phi_{F1}\rangle\langle\Phi_{F2}|\Phi_{2F}\rangle
\biggr].
\end{eqnarray}
The bubble contribution is secured
by plugging the relevant propagators into (\ref{Bmchpt_f0})
\begin{eqnarray}
\label{Bmchpt_p_f0}
B_{2+1,\sigma}^{\rm M\chi PT}(p) &=& \mu^2 \sum_k\Biggl\{
-\frac{4}{3}\frac{1}{(k+p)^2+M_{x,x}^2} \cr
&&\hspace{-1.8cm}
\times\frac{1}{(k^2+M_{x,x}^2)^2}
\frac{(k^2+M_{U_I}^2)(k^2+M_{S_I}^2)}{k^2+M_{\eta_I}^2}\cr
&&\hspace{-1.8cm}
+\frac{4}{9}\frac{1}{((k+p)^2\hspace{-0.1cm}+\hspace{-0.1cm}M_{x,x}^2)^2}
\frac{((k+p)^2\hspace{-0.1cm}+\hspace{-0.1cm}M_{U_I}^2)
((k+p)^2\hspace{-0.1cm}+\hspace{-0.1cm}M_{S_I}^2)}
{(k+p)^2\hspace{-0.1cm}+\hspace{-0.1cm}M_{\eta_I}^2} \cr
&&\hspace{-1.8cm}
\times\frac{1}{(k^2+M_{x,x}^2)^2}
\frac{(k^2+M_{U_I}^2)(k^2+M_{S_I}^2)}{k^2+M_{\eta_I}^2}\cr
&&
\hspace{-1.8cm}+2\frac{1}{(k+p)^2+M_{x,x}^2}\frac{1}{(k^2+M_{x,x}^2)^2} \cr
&&\hspace{-1.8cm}
+2\frac{1}{(k+p)^2+M_{x,u}^2} \frac{1}{k^2+M_{x,u}^2} \cr
&&\hspace{-1.8cm}
+\frac{1}{(k+p)^2+M_{x,s}^2} \frac{1}{k^2+M_{x,s}^2}\Biggl\} .
\end{eqnarray}
It is worth mentioning that $\sigma$ bubble in MA$\chi$PT is much simpler than
that of the corresponding S$\chi$PT result~\cite{Bernard:2007qf},
as expected.
It is convenient to use the partial fraction decomposition,
then expression~(\ref{Bmchpt_p_f0}) can be simplified to a compact form
\begin{eqnarray}
\label{Bmchpt_f0_compact}
B_{2+1,\sigma}^{\rm M\chi PT}(p) &=& B^2 \sum_k\Biggl\{
\frac{h_1}{(k+p)^2+M_{x,x}^2} \frac{1}{k^2+M_{x,x}^2} \cr
&&+\frac{h_2}{(k+p)^2+M_{x,x}^2}    \frac{1}{(k^2+M_{x,x}^2)^2}\cr
&&+\frac{h_3}{(k+p)^2+M_{x,x}^2}    \frac{1}{ k^2+M_{\eta_I}^2}\cr
&&+\frac{h_4}{((k+p)^2+M_{x,x}^2)^2}\frac{1}{(k^2+M_{x,x}^2)^2}\cr
&&+\frac{h_5}{(k+p)^2+M_{\eta_I}^2} \frac{1}{ k^2+M_{\eta_I}^2}\cr
&&+\frac{h_6}{((k+p)^2+M_{x,x}^2)^2}\frac{1}{ k^2+M_{\eta_I}^2}\cr
&&+2\frac{1}{(k+p)^2+M_{x,u}^2} \frac{1}{k^2+M_{x,u}^2} \cr
&&+\frac{1}{(k+p)^2+M_{x,s}^2} \frac{1}{k^2+M_{x,s}^2}\Biggl\} ,
\end{eqnarray}
where
\begin{eqnarray}
\label{Bmchpt_f0_compact_coeff}
h_1 &=& 2-\frac{4}{3}\frac{ 3M_{x,x}^2(M_{x,x}^2-2M_{\eta_I}^2)+2M_{S_I}^4+M_{U_I}^4}
{3(M_{\eta_I}^2-M_{x,x}^2)^2} \cr
&&+\frac{4}{9}\left(\frac{ 3M_{x,x}^2(M_{x,x}^2-2M_{\eta_I}^2)+2M_{S_I}^4+M_{U_I}^4}
{3(M_{\eta_I}^2-M_{x,x}^2)^2}\right)^2, \cr
h_2 &=& \frac{(M_{U_I}^2-M_{x,x}^2)(M_{S_I}^2-M_{x,x}^2)}
{M_{\eta_I}^2-M_{x,x}^2} \times \cr
&&\left(
\frac{8}{9}\frac{ 3M_{x,x}^2(M_{x,x}^2-2M_{\eta_I}^2)+2M_{S_I}^4+M_{U_I}^4}
{3(M_{\eta_I}^2-M_{x,x}^2)^2}-\frac{4}{3} \right),  \cr
h_3 &=& \frac{(M_{U_I}^2-M_{\eta_I}^2)(M_{S_I}^2-M_{\eta_I}^2)}
{(M_{x,x}^2-M_{\eta_I}^2)^2} \times \cr
&&\left(
\frac{8}{9}\frac{ 3M_{x,x}^2(M_{x,x}^2-2M_{\eta_I}^2)+2M_{S_I}^4+M_{U_I}^4}
{3(M_{\eta_I}^2-M_{x,x}^2)^2}-\frac{4}{3} \right),  \cr
h_4 &=& \frac{4}{9}\left(\frac{(M_{U_I}^2-M_{x,x}^2)(M_{S_I}^2-M_{x,x}^2)}
{M_{\eta_I}^2-M_{x,x}^2}\right)^2,  \cr
h_5 &=& \frac{4}{9}\left(\frac{(M_{U_I}^2-M_{\eta_I}^2)(M_{S_I}^2-M_{\eta_I}^2)}
{(M_{x,x}^2-M_{\eta_I}^2)^2}\right)^2,  \cr
h_6 &=& \frac{8}{9}\frac{(M_{U_I}^2-M_{x,x}^2)(M_{S_I}^2-M_{x,x}^2)}
{M_{\eta_I}^2-M_{x,x}^2} \cr
&&\times
\frac{(M_{U_I}^2-M_{\eta_I}^2)(M_{S_I}^2-M_{\eta_I}^2)}
{(M_{x,x}^2-M_{\eta_I}^2)^2}.
\end{eqnarray}

The time Fourier transform of this bubble correlator
(namely
$\displaystyle B_{2+1,\sigma}^{\rm M\chi PT}(t) =
F.T.[B_{2+1,\sigma}^{\rm M\chi PT}(p)]_{\mathbf{p}=\mathbf{0}}
$)
is then provided by
\begin{eqnarray}
\label{eq:bubble_compact_f0_t}
B_{2+1,\sigma}^{\rm M\chi PT}(t) \hspace{-0.15cm}&=&\hspace{-0.15cm}
\frac{\mu^2}{4L^3}\sum_{\mathbf{k}}
\Biggl[
h_1\frac{e^{-2\omega_{xx} t}}{\omega_{xx}^2} +
h_2\frac{e^{-2\omega_{xx}t}}{2\omega_{xx}^4} \left(\omega_{xx} t+1\right) \cr
\hspace{-0.15cm}&&\hspace{-0.15cm}
+h_3\frac{e^{-(\omega_{xx} + \omega_{\eta_I})t}}
{\omega_{xx}\omega_{\eta_I}}
+h_4\frac{e^{-2\omega_{xx}t}}{4\omega_{xx}^6}
\left(\omega_{xx} t + 1\right)^2 \cr
\hspace{-0.15cm}&&\hspace{-0.15cm}
+h_5\frac{e^{-2\omega_{\eta_I} t}}{\omega_{\eta_I}^2}
+h_6\frac{e^{-(\omega_{xx} + \omega_{\eta_I})t}}
{2\omega_{xx}^3\omega_{\eta_I}}\left(\omega_{xx} t + 1\right)    \cr
\hspace{-0.15cm}&&\hspace{-0.15cm}+
  2\frac{e^{-2\omega_{xu}t}}{\omega_{xu}^2}
  +   \frac{e^{-2\omega_{xs}t}}{\omega_{xs}^2}
\Biggr].
\end{eqnarray}

Once again, we note  that no free parameters are presented in (\ref{eq:bubble_compact_f0_t}),
which is solely predicted by MA$\chi$PT.
The expression (\ref{eq:bubble_compact_f0_t})
also gets unphysical contributions from $\pi\eta$ intermediate states,
which luckily never dominate $\sigma$ correlator at large $t$
even if the valence quark masses are enough small.
Moreover, there are two kind of double poles in its momentum-space propagator:
the infrared-sensitive linear-in-$t$ growth factors in the fourth and sixth terms of Eq.~(\ref{eq:bubble_compact_f0_t})arising from the double poles,
and the strong infrared-sensitive quadratic-in-$t^2$ growth factor in the fourth term
of Eq.~(\ref{eq:bubble_compact_f0_t}) stemming from the multiplication of
two double poles.
We will observe the fourth term of Eq.~(\ref{eq:bubble_compact_f0_t})
plays a key role in $\sigma$ bubble.

The unphysical $\pi\eta$ intermediate states contribute to  $\sigma$ bubble (\ref{eq:bubble_compact_f0_t})
since MA$\chi$PT is not unitary at $a=0$.
The unphysical $\pi\eta$ contributions can not be neatly removed for
any selection of mixed-action realistic simulation parameters.
In  the continuum limit ($a^2\Delta_I\to 0$, $a^2\Delta_{\rm Mix}\to 0$),
the expression (\ref{eq:bubble_compact_f0_t})
reduces to a pretty simple form
\begin{eqnarray}
\label{eq:bubble_compact_limit_f0_t}
\hspace{-1.5cm} B_{2+1,\sigma}^{\rm M\chi PT}(t) &=&
\frac{\mu^2}{4L^3}\sum_{\mathbf{k}}
\Biggl[
3\frac{e^{-2\omega_{U_5}t}}{\omega_{U_5}^2} +
 \frac{e^{-2\omega_{K_5}t}}{\omega_{K_5}^2}\cr
&&+\frac{1}{9}
\frac{e^{-2\omega_{eta_5}t}}{\omega_{\eta_5}^2}
\Biggr],
\end{eqnarray}
which is elegantly consistent with the corresponding S$\chi$PT result,
which is courteously written down in (\ref{BSchpt_limit_f0}).
Without doubt, away from the continuum limit,
the wanted physical contributions $e^{- 2M_\pi t}$
will dominate $\sigma$ bubble at large $t$.

\section{Numerical illustration of bubble contributions}
\label{sec:Num}
We plan on launching a serial of the mixed action lattice investigations
of scalar mesons using DW fermions
on top of the MILC $2+1$ asqtad-improved staggered sea quarks.
So far, the MILC lattice ensembles with two lattice spacings:
the MILC coarse lattices and MILC fine lattices are extensively studied.
Therefore, it is very useful to exploit the known
parameters determined by MILC collaboration on these lattices~\cite{Bernard:2001av}
to acquire the preliminary numerical predictions
for the bubble contributions to scalar mesons
in MA$\chi$PT, which then will guide us
in the ongoing lattice investigations with the reasonable and
economical mass tuning between the valence quarks and sea quarks.

We illustrate these predictions only on two MILC lattice ensembles:
one is a coarse ensemble ($a \approx 0.12$~fm, $am_u/am_s=0.005/0.05$),
another one is a fine ensemble ($a \approx 0.09$~fm, $am_u/am_s=0.0062/0.031$),
which are in this work labeled ``coarse" and ``fine" lattice ensemble, respectively.
The taste-singlet breaking $a^2\Delta_I$ of
sea-sea $\pi$ mass is acquired from MILC determinations~\cite{Bernard:2001av},
and taste breaking parameter $a^2\Delta_{\rm Mix}$
of  valence-sea $\pi$ mass is quoted from Aubin {\it et al} 's measurements~\cite{Aubin:2008wk}.

Considering that MA$\chi$PT violates unitarity at $a\neq0$,
in principle, we can make a huge of choices, and
can not prefer one mass matching to another for a priori reason.
Nonetheless, all tuning choices are identical at $a=0$.
Therefore, we just exhibit two most popular mass matchings
of chiral valence quarks and staggered sea quarks.
To help one quantitatively comprehend each term in the bubble contributions
to scalar mesons, each of them is displayed in the corresponding figures,
and indicate the whole bubble contribution with black solid line.

\subsection{Matching 1}
The first selection is to fix the valence pion mass and kaon mass to
be equal to the taste-pseudoscalar sea pion mass and kaon mass,
to be specific, $M_{x,x}=M_{U_5}$, which is practiced
in Ref.~\cite{mixed_lat}, and $M_{x,z}=M_{K_5}$.
This tuning is attractive since the taste-pseudoscalar pion mass
disappears in the chiral limit, even at $a\ne0$.
Nonetheless, this choice increases unitarity-violation for
the bubble contributions to scalar mesons
due to large taste-singlet breaking $a^2\Delta_I$ on the coarse MILC lattices.

\begin{figure}[h!]
\includegraphics[width=8.0cm,clip]{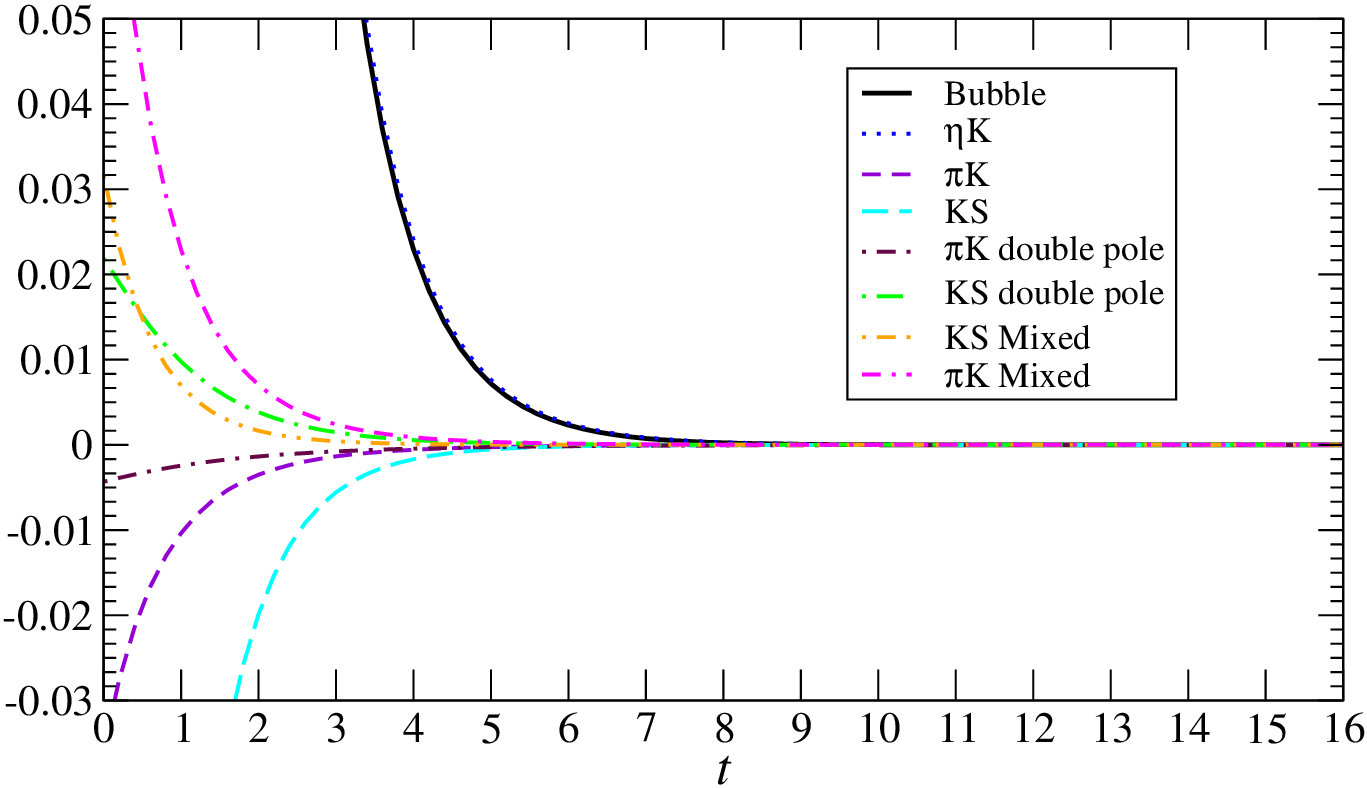}
\includegraphics[width=8.0cm,clip]{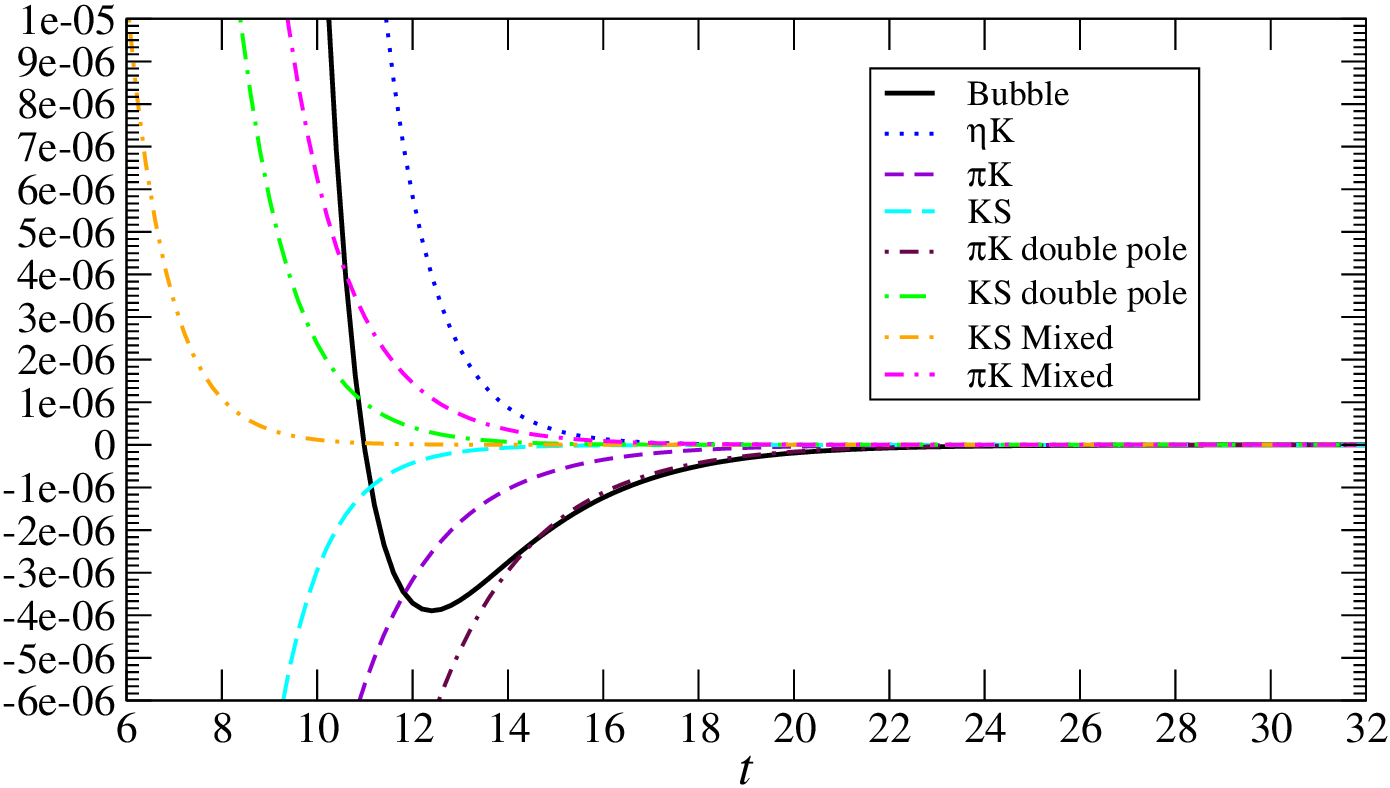}
\caption{\label{fig.mchpt_tune1_kappa_005}
$\kappa$ bubble (\ref{eq:bubble_compact_kappa_t})
for the simulation with chiral fermions
on a MILC ``coarse" staggered configuration.
The valence and sea quark masses are tuned
by matching $M_{x,x}=M_{U_5}$ and $M_{x,z}=M_{K_5}$.
The parameters $a^2\Delta_I$ and $a^2\Delta_{\rm Mix}$ are
taken from Refs.~\cite{Bernard:2001av},~\cite{Aubin:2008wk}, respectively.
The top panel shows the data on the small $t$ range,
while bottom panel indicates the data on the large $t$ range.
}
\end{figure}

\begin{figure}[h!]
\includegraphics[width=8.0cm,clip]{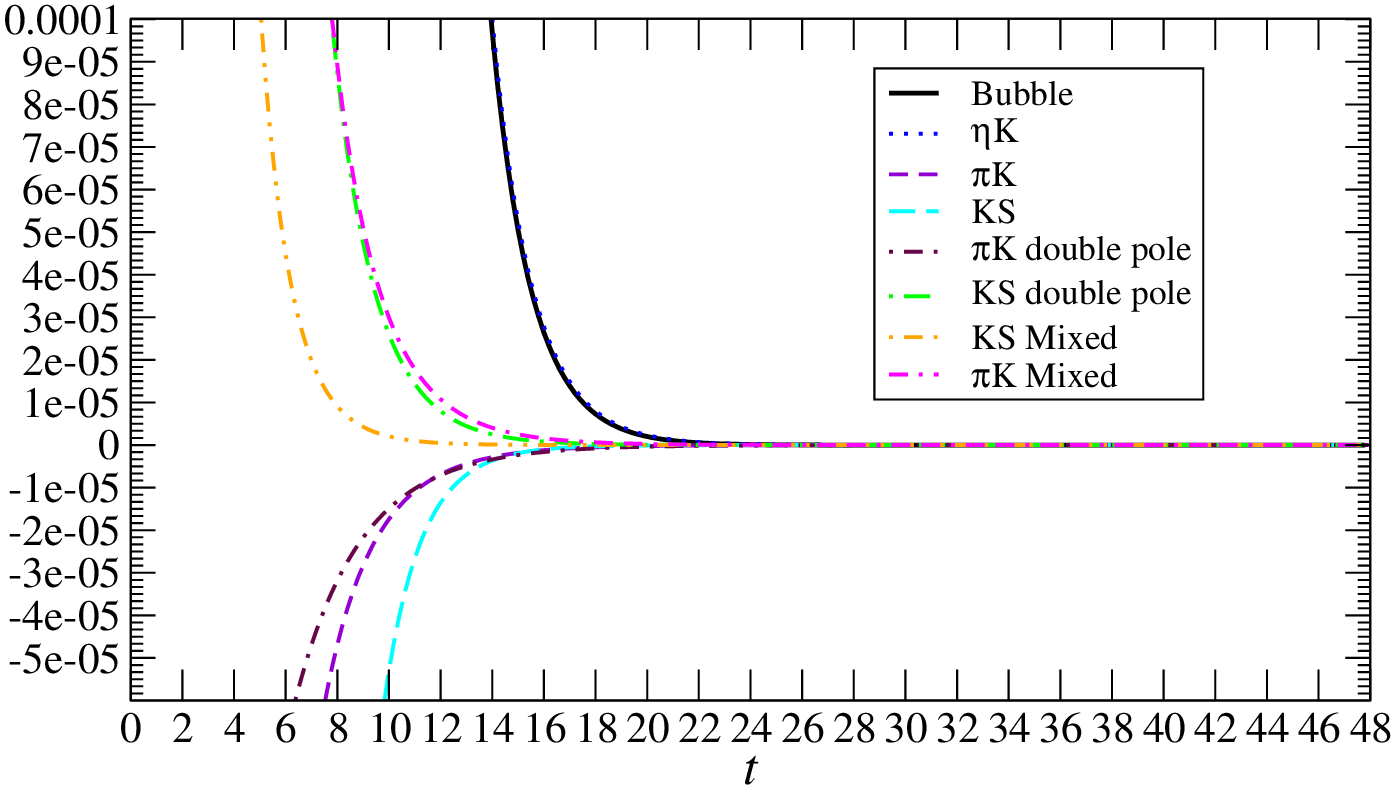}
\includegraphics[width=8.0cm,clip]{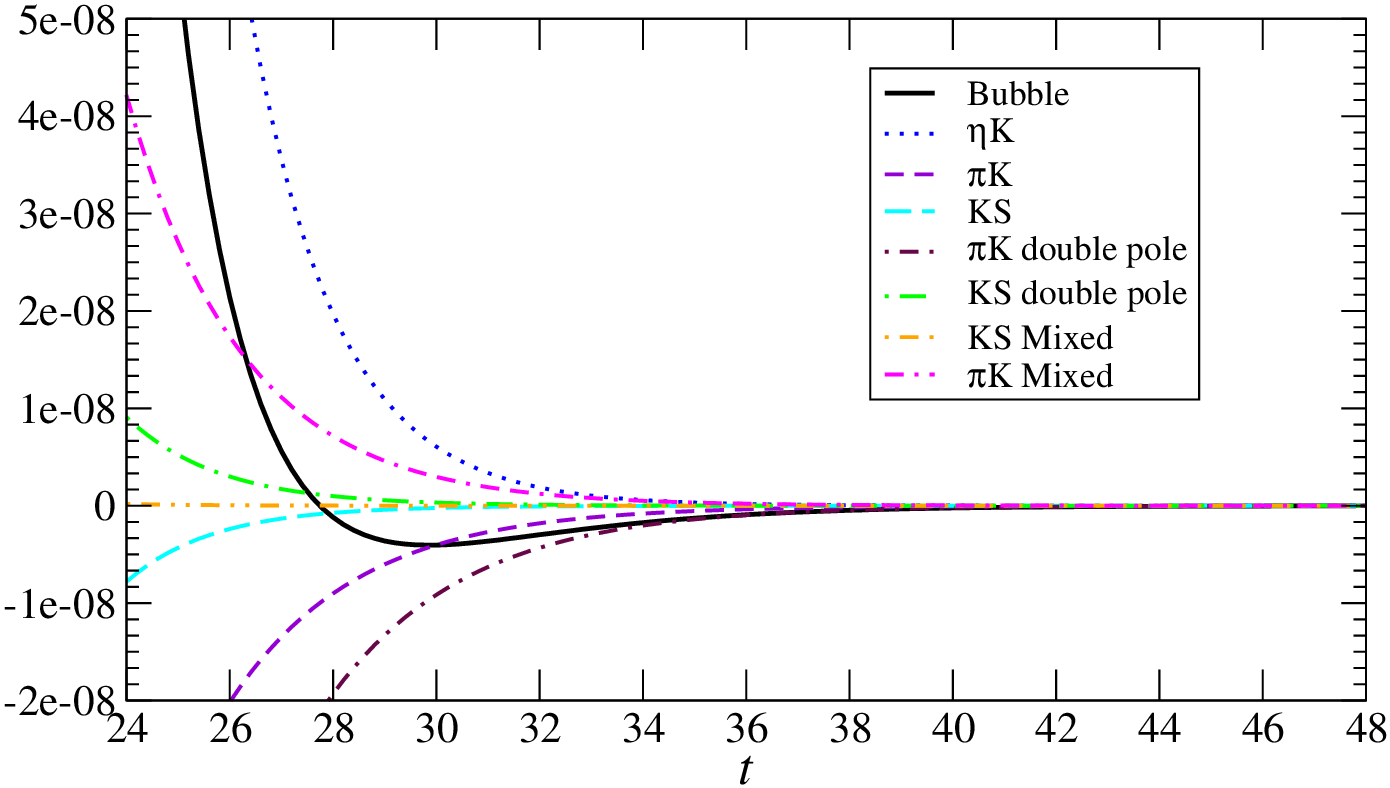}
\caption{\label{fig.mchpt_tune1_kappa_0062}
Same as Fig.~\ref{fig.mchpt_tune1_kappa_005}, but with chiral fermions
on a MILC staggered fine lattice ensemble.
}
\end{figure}

Figures~\ref{fig.mchpt_tune1_kappa_005} and~\ref{fig.mchpt_tune1_kappa_0062}
show $\kappa$ bubble on the MILC coarse and fine ensembles, respectively.
In these figures, ``$\pi K$" indicates the intermediate states
with the valence pion and kaon, and likewise for ``$K\eta$" and ``$KS$",
while ``$\pi K$ Mixed"  represents
the intermediate states with the mixed valence-sea pion and kaon,
and likewise for ``$KS$ Mixed".
``$\pi K$ double pole" and ``$K S$ double pole" are the fourth and fifth terms
in (\ref{eq:bubble_compact_f0_t}).
The analogous notations are employed to $\sigma$ and $a_0$ bubbles in the
corresponding figures in this work.


From Figs.~\ref{fig.mchpt_tune1_kappa_005} and~\ref{fig.mchpt_tune1_kappa_0062},
we note that physical $\eta K$ states dominate $\kappa$ bubble
until $t\approx11$ for coarse lattice and $t\approx27$ for fine lattice
and quickly decrease after that.
On the other hand,  ``$\pi K$ double pole" is pretty small at small $t$,
but for enough large $t$ it gradually dominates $\kappa$ bubble,
whereas ``$KS$ double pole" is negligible,
and it is important to note that there is a cancelation
between two double pole terms, which is a special feature of the $\kappa$ bubble,
and this cancelation  is a good news for studying $\kappa$ meson
since it decreases the unitarity-violation of $\kappa$ bubble.
Moreover, the $\pi K$ state plays an important in $\kappa$ bubble,
and it is important to note that unphysical $KS$ state never
dominate $\kappa$ bubble.

\begin{figure}[h!]
\includegraphics[width=8.0cm,clip]{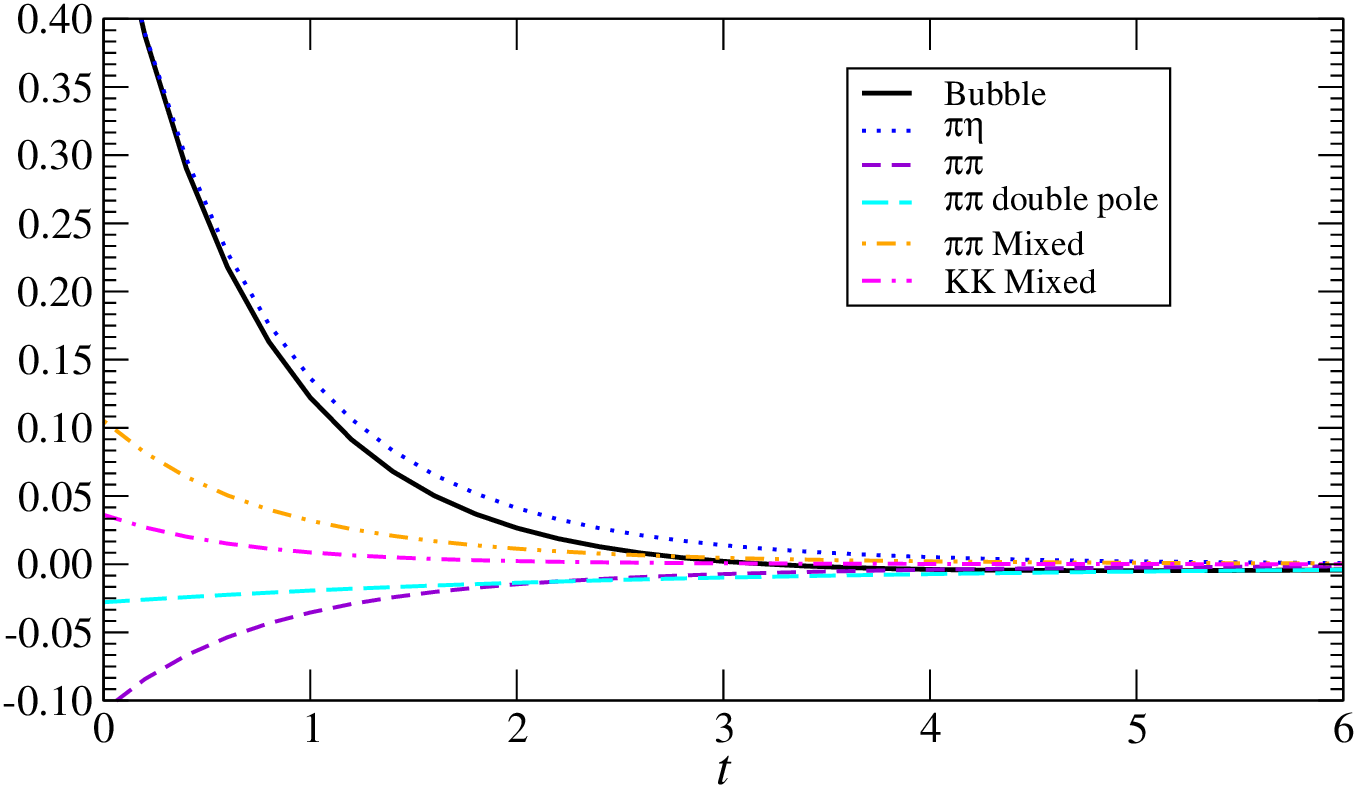}
\includegraphics[width=8.0cm,clip]{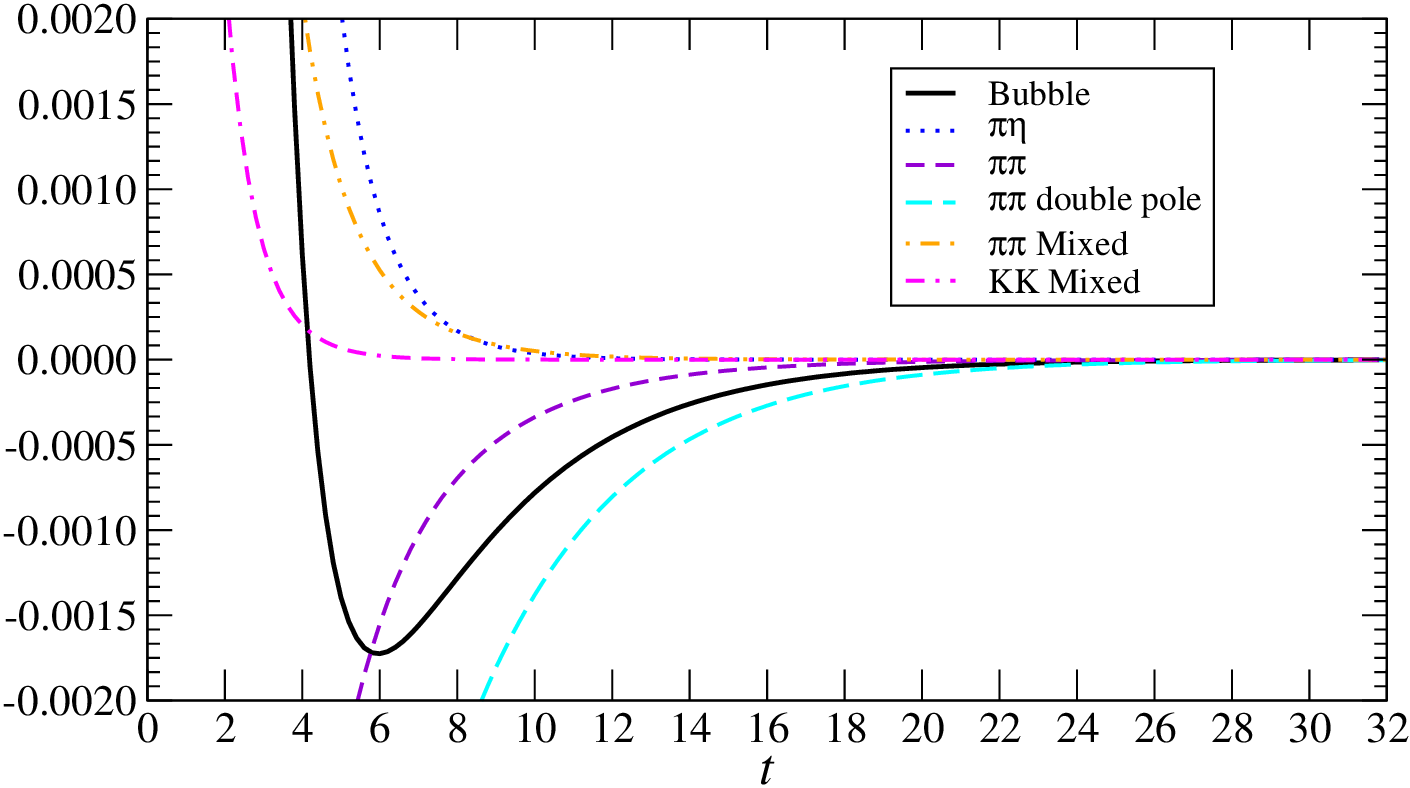}
\caption{\label{fig.mchpt_tune1_a0_005}
$a_0$ bubble (Eq.~(11) of Ref.~\cite{Aubin:2008wk})
for a simulation with chiral fermions
on a MILC coarse staggered configuration.
The valence and sea quark masses are tuned as $M_{x,x}=M_{U_5}$.
The top panel shows the data on the small $t$ range,
while bottom panel indicates the data on the large $t$ range.
``$\pi\pi$ double pole" is the third term in Eq.~(11) of Ref.~\cite{Aubin:2008wk}.
}
\end{figure}

\begin{figure}[h!]
\includegraphics[width=8.0cm,clip]{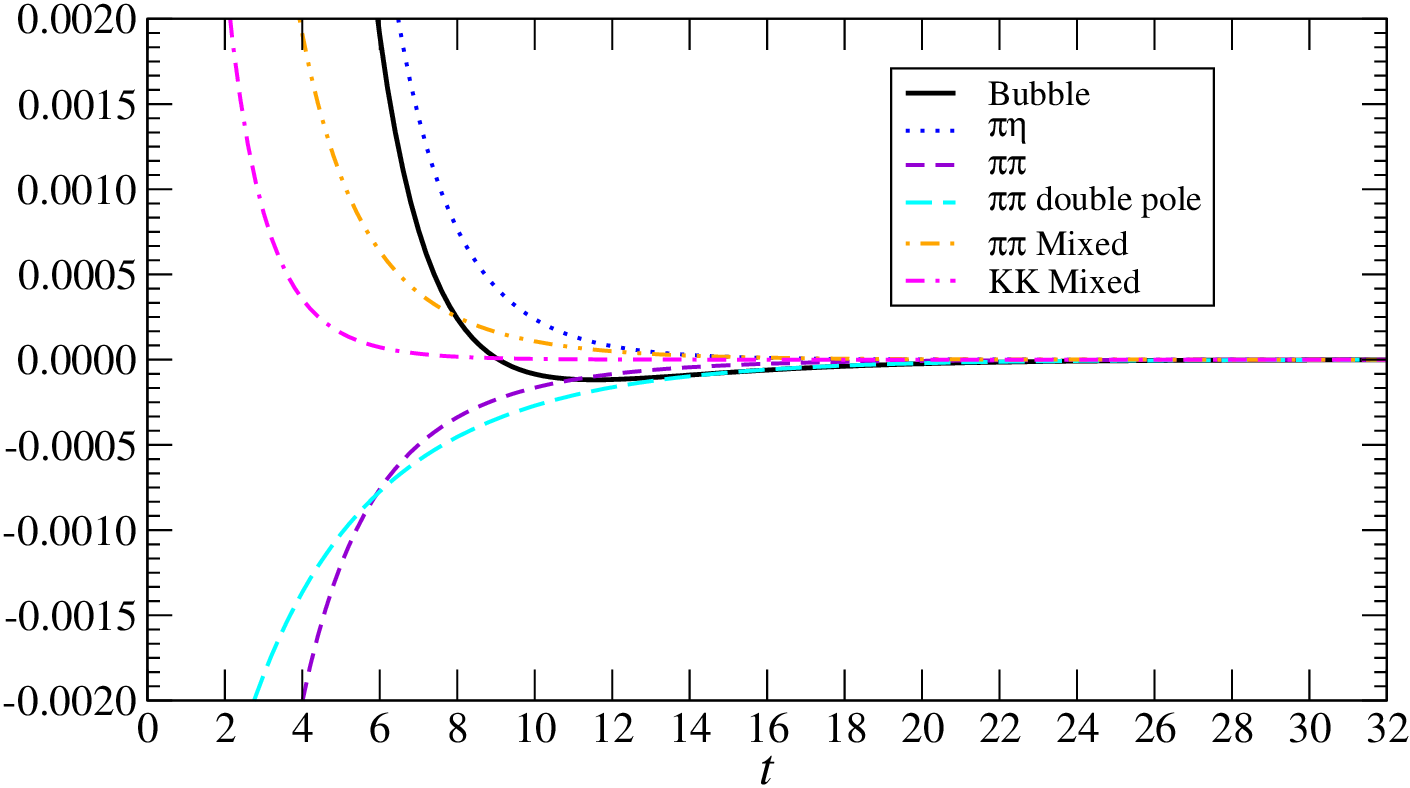}
\includegraphics[width=8.0cm,clip]{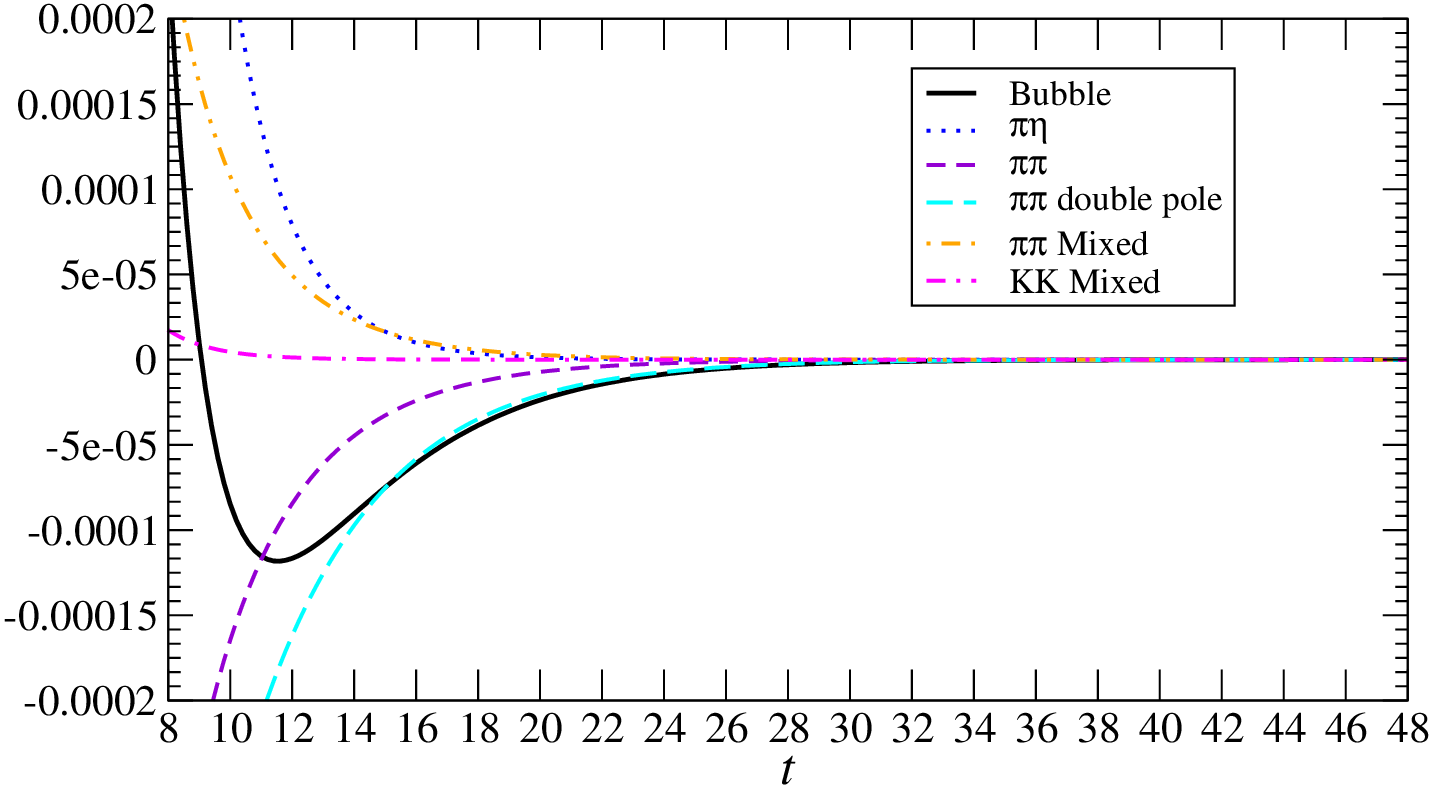}
\caption{\label{fig.mchpt_tune1_a0_0062}
Same as Fig.~\ref{fig.mchpt_tune1_a0_005}, but with chiral fermions
on a MILC staggered fine lattice ensemble.
}
\end{figure}

Figures~\ref{fig.mchpt_tune1_a0_005} and~\ref{fig.mchpt_tune1_a0_0062}
show $a_0$ bubble on the MILC coarse and fine ensembles, respectively.
We note that physical $\pi\eta$ states
dominate $a_0$ bubble until $t\approx4$ for coarse lattice
and $t\approx12$ for fine lattice,
and quickly decrease after that.
On the other hand, the third term in Eq.~(11) of Ref.~\cite{Aubin:2008wk}
(``$\pi\pi$ double pole") is pretty small at small $t$,
but for enough large $t$ it eventually dominates $a_0$ bubble.
It is important to note that physical $KK$ state never
dominate $a_0$ bubble, while unphysical $\pi\pi$ state
plays a very important role in $a_0$ bubble.
This indicates that the reliable determination of $a_0$ meson mass
is feasible only for appropriate quark masses and times~\cite{Prelovsek:2005rf}.

\begin{figure}[h!]
\includegraphics[width=8.0cm,clip]{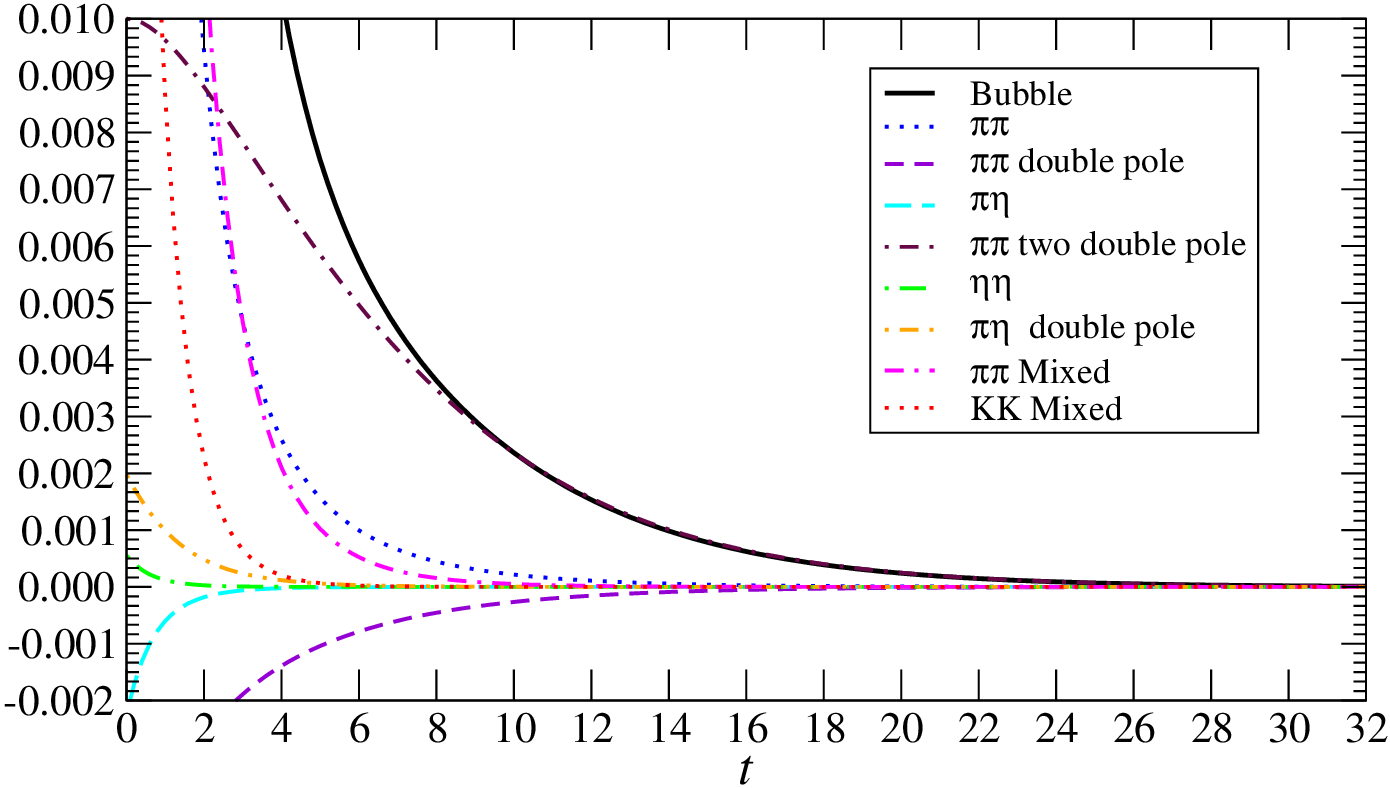}
\includegraphics[width=8.0cm,clip]{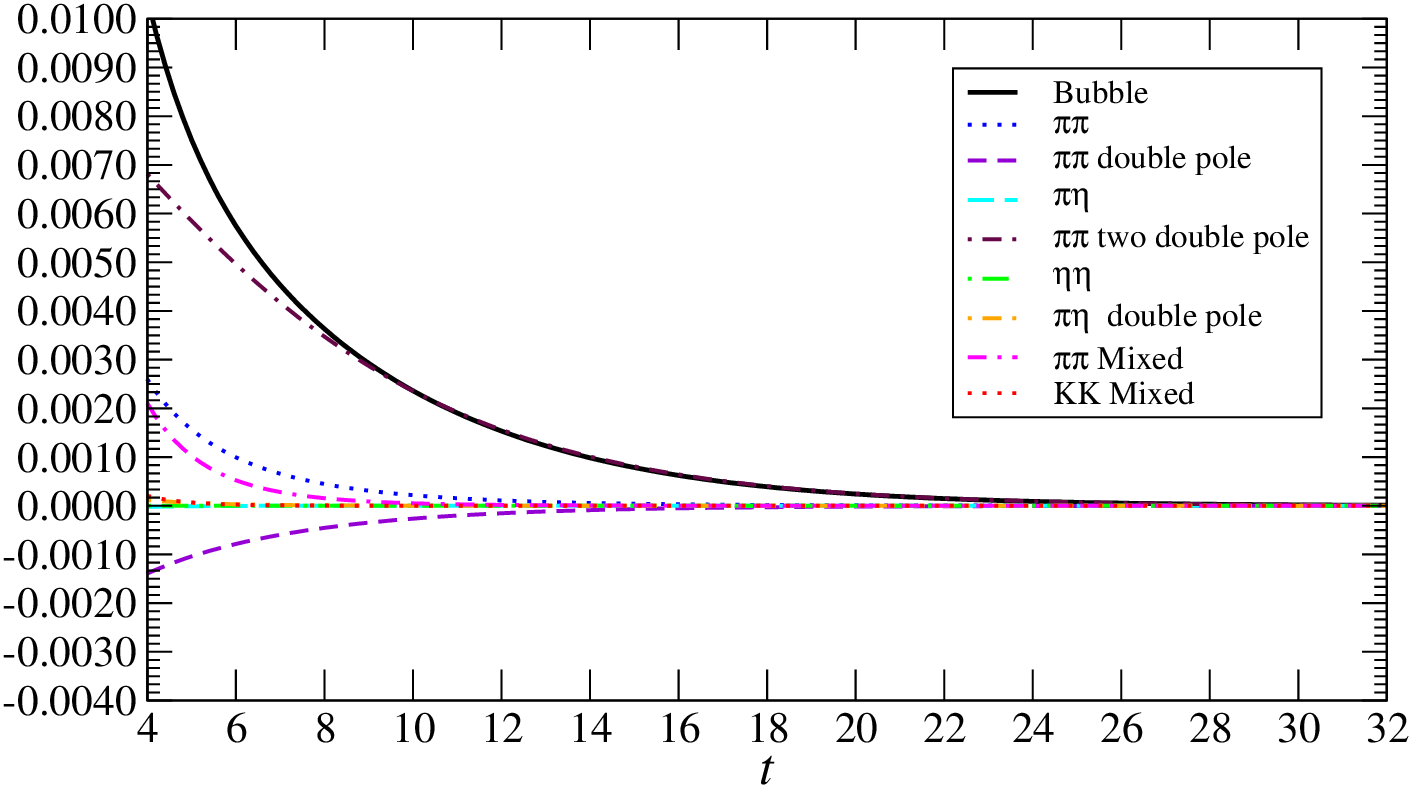}
\caption{\label{fig.mchpt_tune1_f0_005}
$\sigma$ bubble (\ref{eq:bubble_compact_f0_t})
for a simulation with chiral fermions
on a MILC coarse staggered configuration.
The valence and sea quark masses are tuned as $M_{x,x}=M_{U_5}$.
}
\end{figure}

\begin{figure}[h!]
\includegraphics[width=8.0cm,clip]{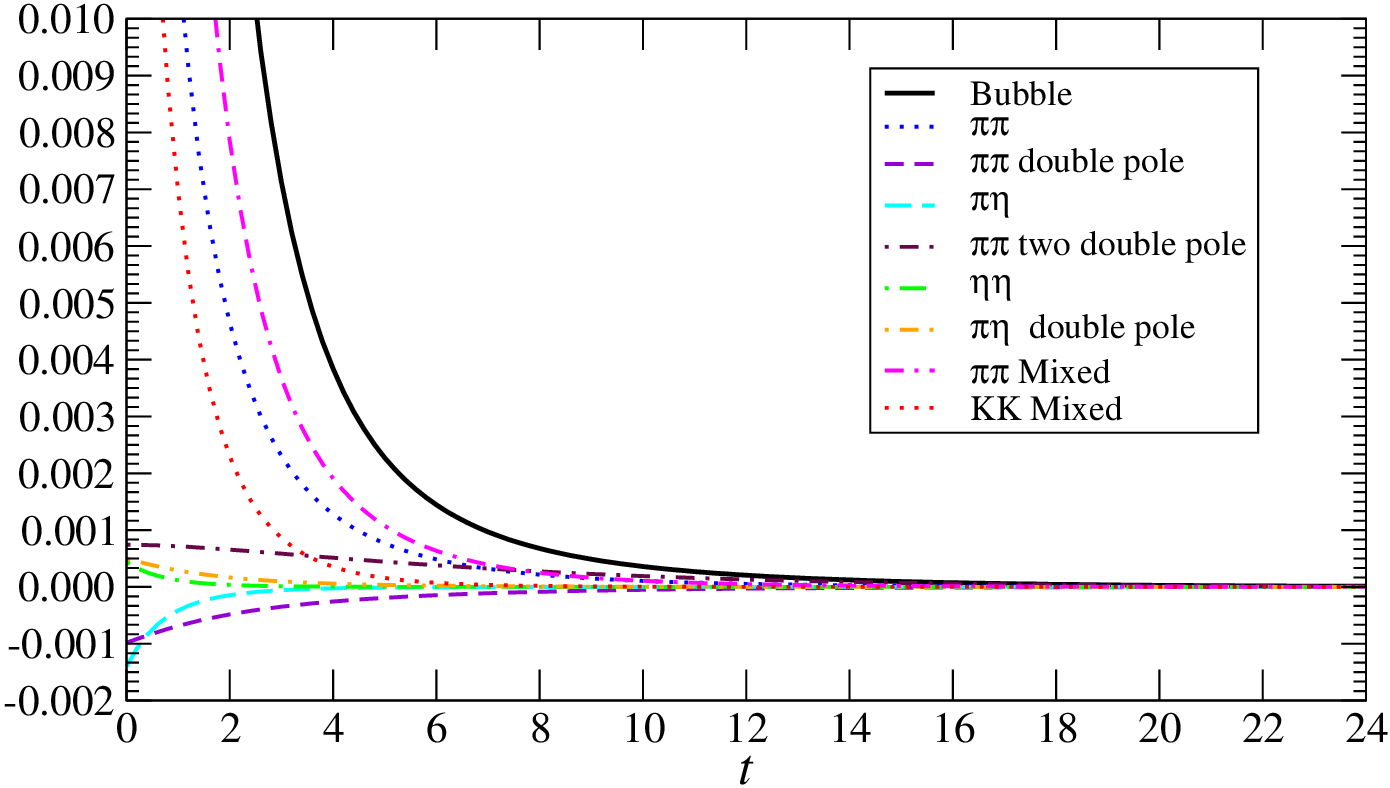}
\includegraphics[width=8.0cm,clip]{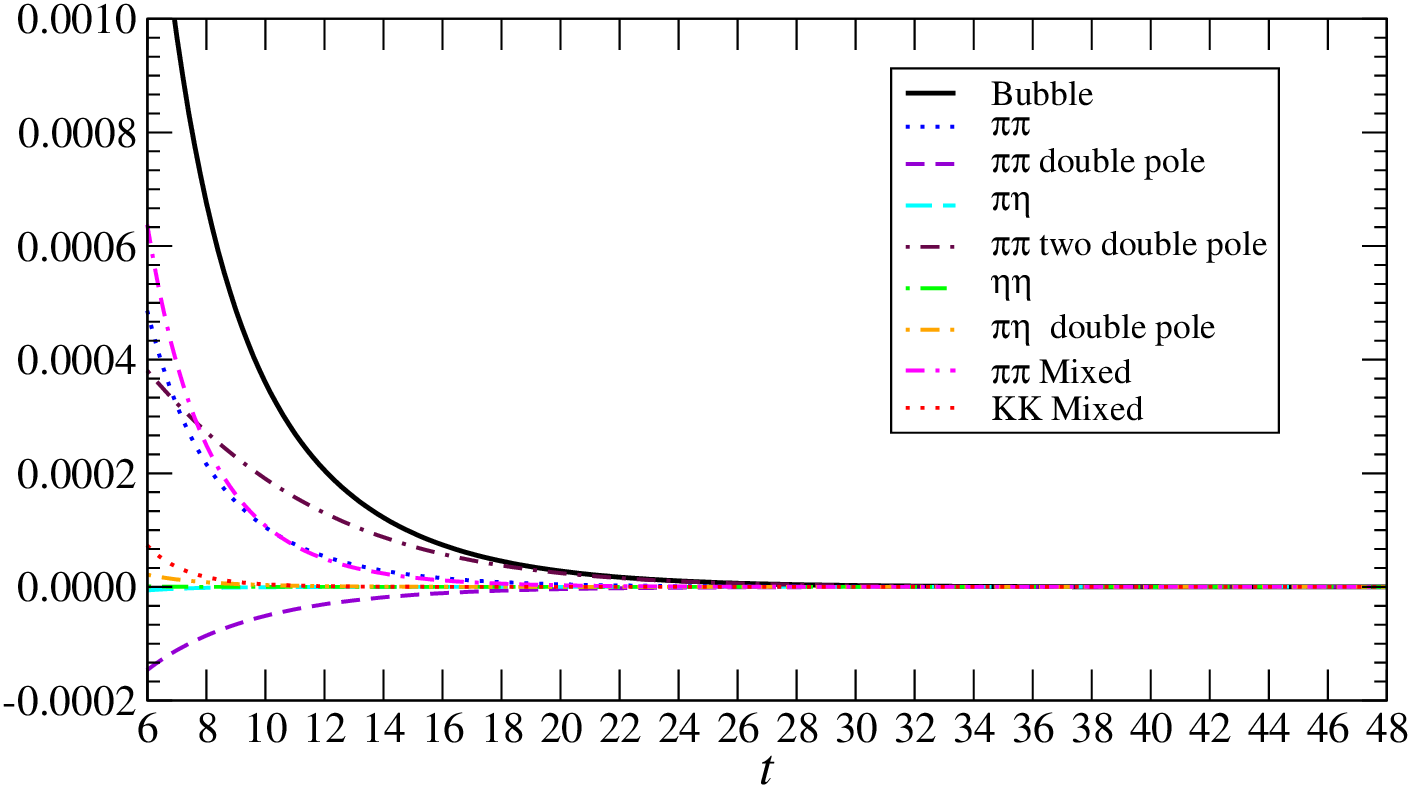}
\caption{\label{fig.mchpt_tune1_f0_0062}
Same as Fig.~\ref{fig.mchpt_tune1_f0_005}, but with chiral fermions
on a MILC staggered fine lattice ensemble.
}
\end{figure}

\clearpage
\newpage
Figures~\ref{fig.mchpt_tune1_f0_005} and~\ref{fig.mchpt_tune1_f0_0062}
show $\sigma$ bubble on the MILC coarse and fine ensembles, respectively.
We note that physical $\pi\pi$ states (the second term in (\ref{eq:bubble_compact_f0_t}))
dominates $\sigma$ bubble until $t\approx3$ for coarse lattice
and $t\approx8$ for fine lattice.
On the other hand,  the fourth term in (\ref{eq:bubble_compact_f0_t})
(``$\pi \pi$ two double pole") is pretty small at small $t$~\footnote{
The coefficient of the factor quadratic-in-$t^2$, which is denoted
as $h_4$ in~(\ref{Bmchpt_f0_compact_coeff}),
is quite small for a typical lattice simulation~\cite{Golterman:2005xa}.
It is about $0.005$ for coarse lattice and $0.0007$ for fine lattice.
},
but for enough large $t$ it eventually dominates $\sigma$ bubble~\footnote{
It overwhelmingly dominates $\sigma$ bubble as early as  $t\approx6$ for coarse lattice,
and dominates $\sigma$ bubble about $t\approx16$  for fine lattice.
It means that if we do not choose a suitable simulation parameters,
this term will be ``infrared-diaster" to $\sigma$ bubble.
}.
While the sixth term in (\ref{eq:bubble_compact_f0_t})
(``$\pi\eta$ double pole") is negligible, and it is obvious to note that
there exists a cancelation among three term with double poles.
Moreover, it is important to note that the unphysical $\pi\eta$ state never
dominate $\sigma$ bubble,
and $\sigma$ bubble is positive for all time.

It is worth mentioning that all the bubble contributions to scalar correlators
are dominated by the double poles at large $t$ for this tuning.
Actually the double pole is an unphysical effect
which stems from selecting the valence-quark action different from
the sea-quark action~\cite{Golterman:2005xa}.
We found that these unitarity violations are not likely to be fairly small
for this tuning on MILC coarse lattice,
while the degree of unitarity violation decreases with the finer lattice spacing, as we expected.

\subsection{Matching 2}
The second choice is to fix the valence pion mass and valence kaon mass
to be equal to the taste-singlet sea pion mass $U_I$ and taste-singlet sea kaon mass,
respectively. To be specific, $M_{x,x}=M_{U_I}$ and $M_{x,z}=M_{K_I}$.
This tuning one hundred per cent removes the double pole terms in
both $\kappa$ bubble in (\ref{eq:bubble_compact_kappa_t}),
$\sigma$ bubble in (\ref{eq:bubble_compact_f0_t}),
and $a_0$ bubble in Eq.~(11) of Ref.~\cite{Aubin:2008wk}.
Moreover, these bubble contributions are turned out to
be relatively small and usually positive.
Nonetheless it still does not entirely remove unphysical
$KS$ state in $\kappa$ bubble in (\ref{eq:bubble_compact_kappa_t}),
$\pi\eta$ state in $\sigma$ bubble in (\ref{eq:bubble_compact_f0_t}),
and $\pi\pi$ state in $a_0$ bubble in  Eq.~(11) of Ref.~\cite{Aubin:2008wk}.
Additionally, in practice this tuning may not be recommendable
since it would lead to a fairly heavy valence
pion and kaon on the MILC coarse lattices,
as for the MILC fine, superfine, or ultrafine lattices, it is a different story.

\begin{figure}[h!]
\includegraphics[width=8.0cm,clip]{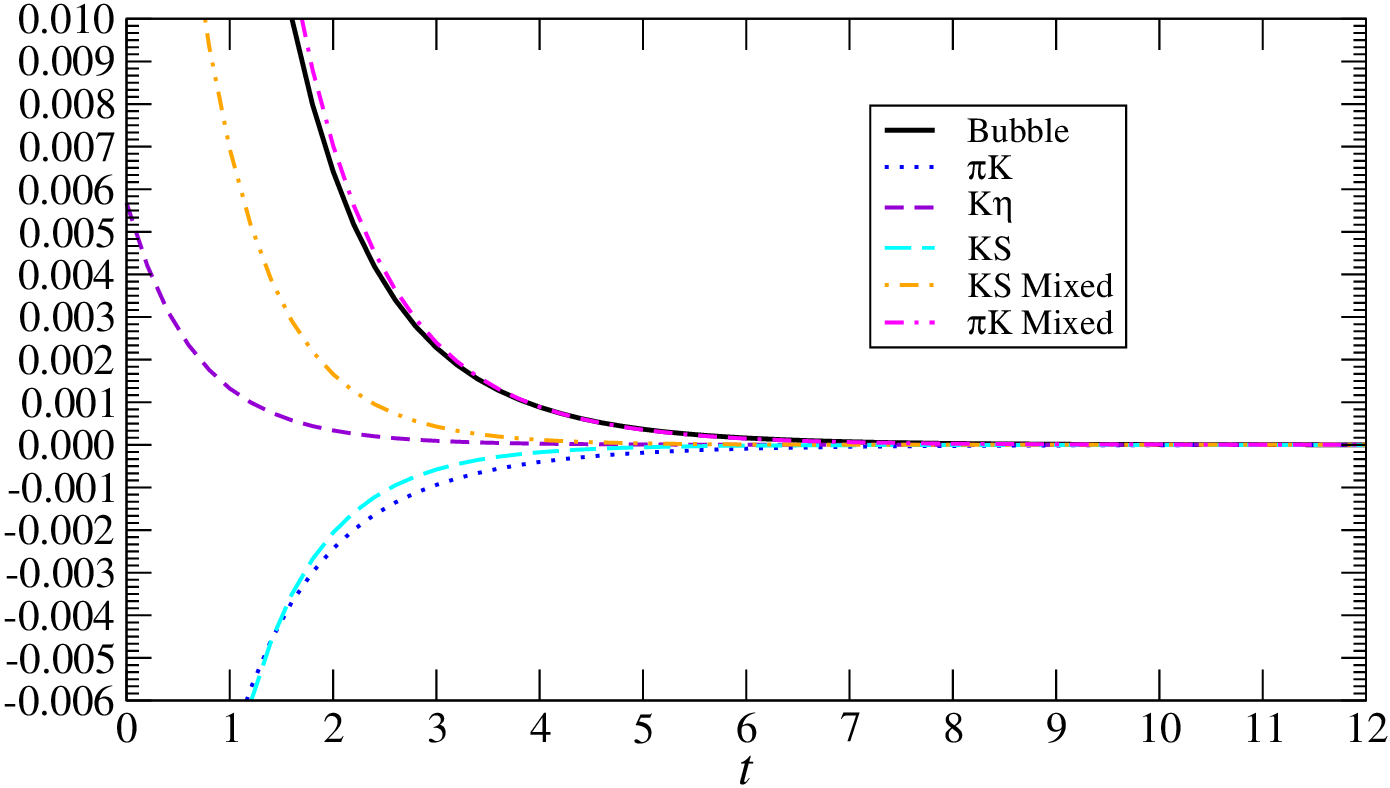}
\includegraphics[width=8.0cm,clip]{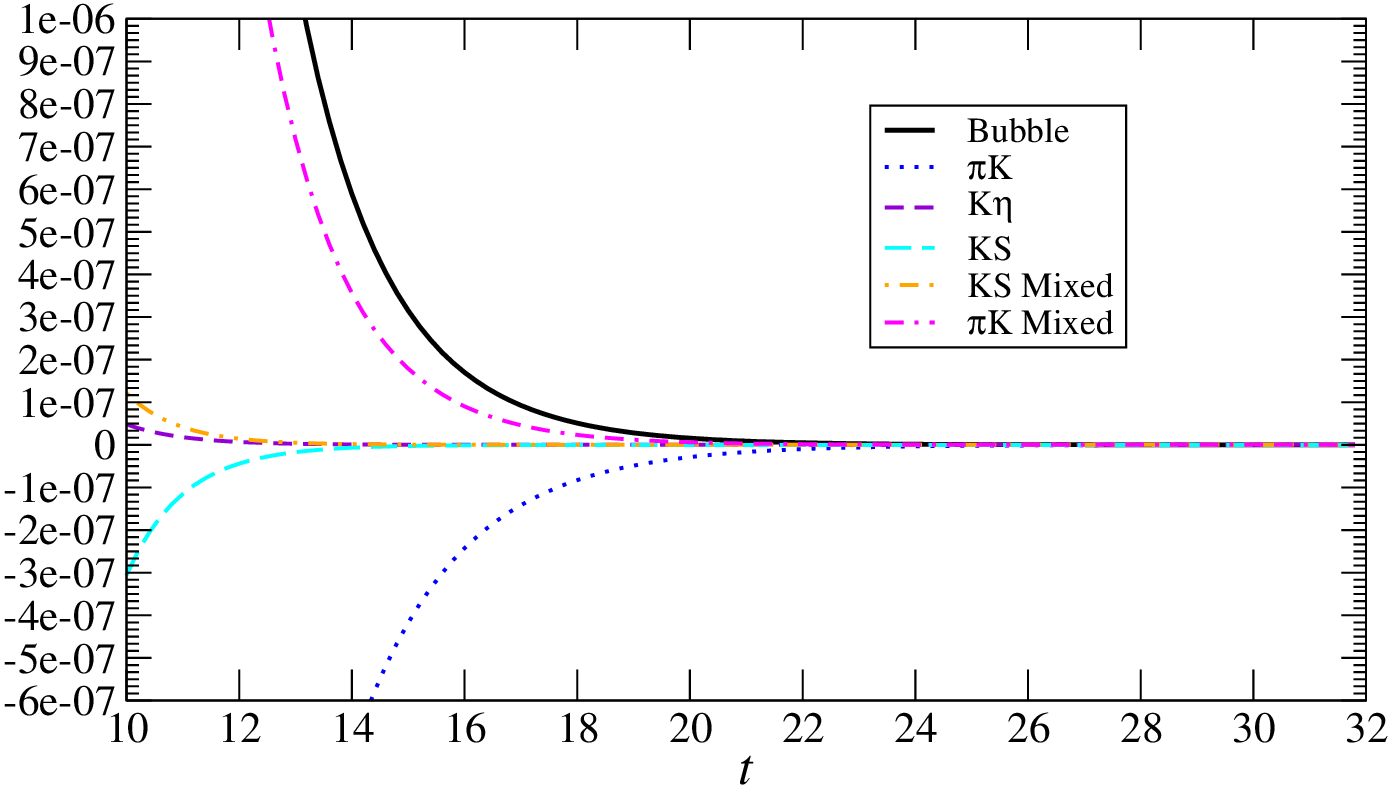}
\caption{\label{fig.mchpt_tune2_kappa_005}
$\kappa$ bubble (\ref{eq:bubble_compact_kappa_t})
for the simulation with chiral fermions
on $2+1$ MILC coarse staggered configuration.
The valence and sea quark masses are tuned
by matching $M_{x,x}=M_{U_I}$ and $M_{x,z}=M_{K_I}$.
}
\end{figure}

\begin{figure}[h!]
\includegraphics[width=8.0cm,clip]{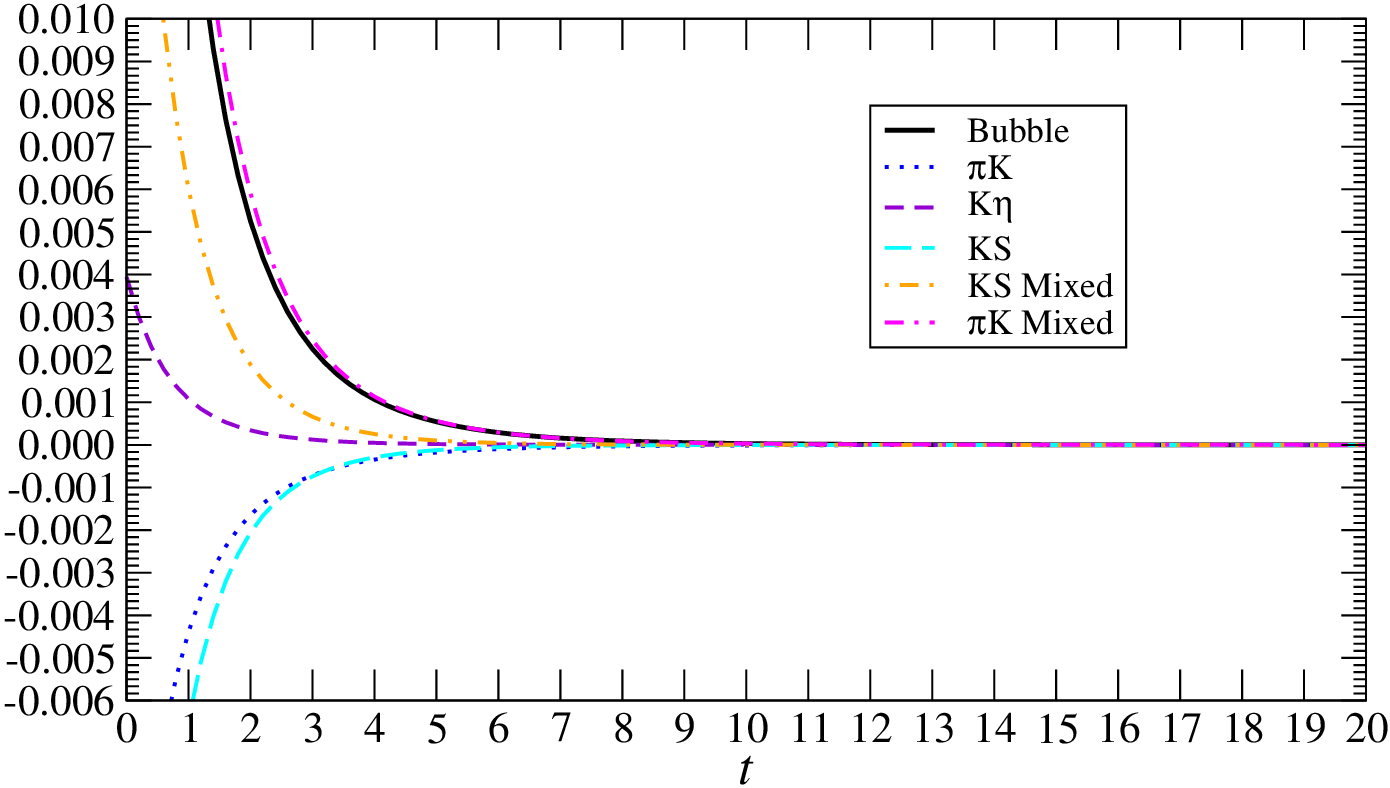}
\includegraphics[width=8.0cm,clip]{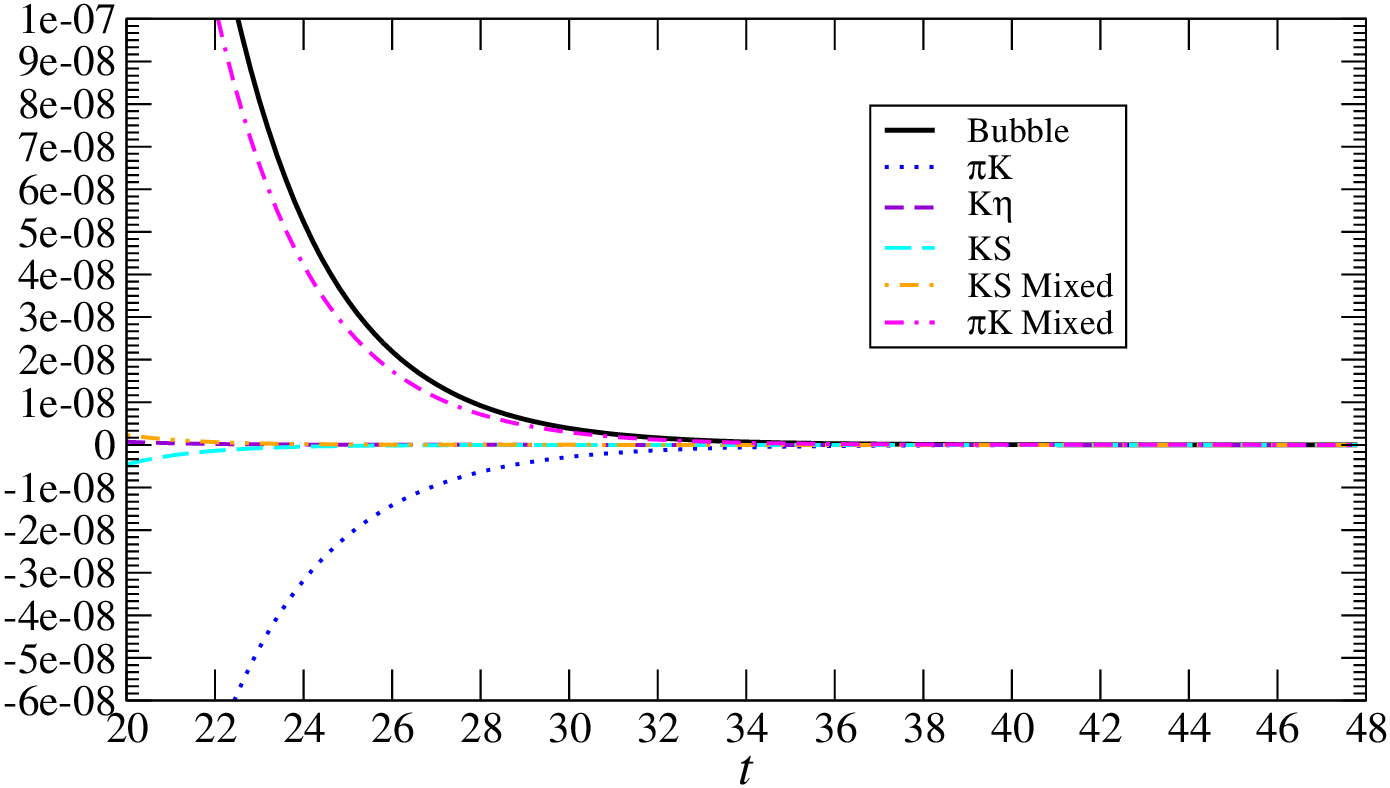}
\caption{\label{fig.mchpt_tune2_kappa_0062}
Same as Fig.~\ref{fig.mchpt_tune2_kappa_005}, but with chiral fermions
on a MILC staggered fine lattice ensemble.
}
\end{figure}

Figure~\ref{fig.mchpt_tune2_kappa_005} and Figure~\ref{fig.mchpt_tune2_kappa_0062}
show $\kappa$ bubble on a MILC coarse and fine ensemble, respectively.
We note that
physical $\pi K$ states (the seventh term in (\ref{eq:bubble_compact_kappa_t}))
dominate $\kappa$ bubble at all time.
Moreover, the unphysical $K S$ state is small,
and never dominate $\kappa$ bubble,
while physical $K\eta$ state make a small contribution to $\kappa$ bubble.

\begin{figure}[h!]
\includegraphics[width=8.0cm,clip]{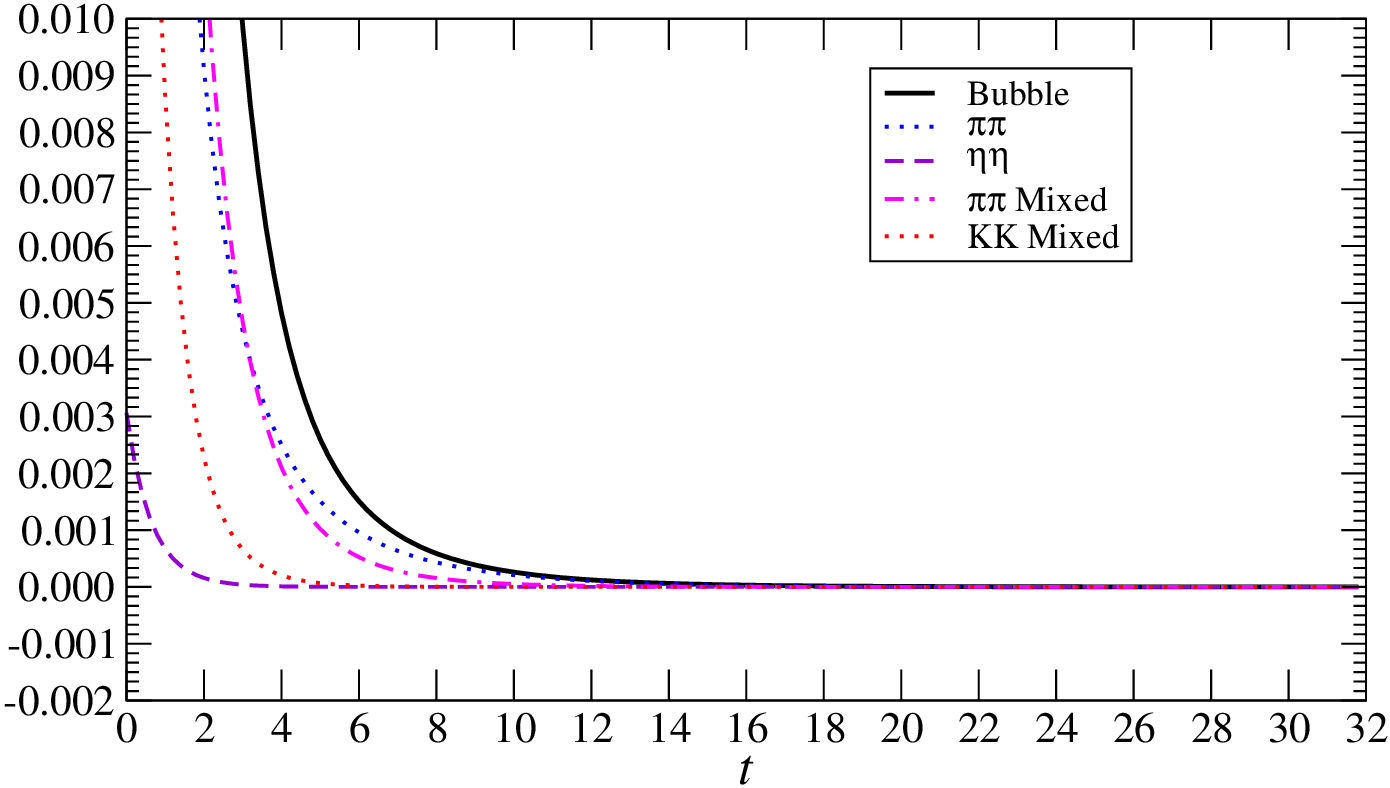}
\caption{\label{fig.mchpt_tune2_f0_005}
$\sigma$ bubble (\ref{eq:bubble_compact_f0_t})
for the simulation with chiral fermions
on $2+1$ MILC fine staggered configuration.
The valence and sea quark masses are tuned
as $M_{x,x}=M_{U_I}$.
}
\end{figure}

Figure~\ref{fig.mchpt_tune2_f0_005} displays
$\sigma$ bubble on a MILC coarse ensemble.
We note that physical $\pi\pi$ states
(the first term in (\ref{eq:bubble_compact_f0_t}))
dominate $\sigma$ bubble at all time.
Moreover, physical $\eta\eta$ and $K K$ states play a small role in $\sigma$ bubble.

\section{Summary}
\label{Sec:Conclusions}
In this work we extended Sasa's derivation on bubble contribution
to $a_0$ correlator in MA$\chi$PT~\cite{Prelovsek:2005rf}
to those of $\kappa$ and $\sigma$ correlators.
We found that these extensions are useful
since $\kappa$ and $\sigma$ bubbles
demonstrate many new features as compared with $a_0$ bubble.
For example, the $\kappa$ and $\sigma$ bubbles are
dominated by the physical two-particle states at enough large $t$,
while $a_0$ bubble is dominated by unphysical two pions states.
Moreover, we notice a strong infrared-sensitive quadratic-in-$t^2$ growth factor
in the $\sigma$ bubble due to the multiplication of two double poles,
in practice, special attention should be paid to monitor the size of this unitarity violations,
otherwise it leads to an ``infrared-diaster" to $\sigma$ bubble for a given tuning.

MA$\chi$PT predicts the observed unitarity-violations pretty well,
it is good news for one who employs the mixed actions to study scalar meson.
Moreover, different mass tunings have remarkable effects on scalar correlators,
our extended analytical expressions are helpful to aid researchers
in selecting the appropriate simulation parameters
for determining scalar meson masses.

Lattice studies of scalar mesons using DW valence quarks and staggered sea quarks
contains the excellent traits of both fermion discretizations.
The corresponding numerical results will be used as a cross-check 
with our lattice studies on scalar mesons
with other methods~\cite{Fu:2012gf}.
Therefore, it will be interesting to measure the relevant lattice data
to verify the MA$\chi$PT formulae for $\kappa$ and $\sigma$ bubbles.
We will appeal for computational resources
to pursue this challenging enterprise.

\section*{Acknowledgments}
We appreciate K. F. Liu for helping us with  some knowledge about scalar meson,
and bring us attention to domain-wall fermion.
We benefit from the happy day of studying scalar mesons with
Carleton DeTar, Claude Bernard and  Sasa Prelovsek during my Ph.D.

\appendix
\section{Bubble contribution}
\label{sec_bubble}
The $\kappa$ bubble term is~\cite{Fu:2011xb,Fu:2013sua}
\begin{eqnarray}
B_{2+1,\kappa}^{\rm S\chi PT}(t) \hspace{-0.2cm}& =&\hspace{-0.2cm}
\mu^2/(4L^3) \times \left\{ f_{B}^\kappa(t) + f_{V}^\kappa(t) + f_{A}^\kappa(t) \right\}, \cr
f_{V}^\kappa(t)  \hspace{-0.2cm}&\equiv& \hspace{-0.2cm}\sum_{\bf k} \frac{1}{2} \bigg\{
g_{V_{\eta }} \frac{e^{-(\omega_{K_V} + \omega_{\eta_V}) t}}{\omega_{K_V}\omega_{\eta_V}} +
g_{V_{\eta^\prime }} \frac{e^{-(\omega_{K_V} + \omega_{\eta'_V}) t}}{\omega_{K_V}\omega_{\eta'_V}}
\cr
\hspace{-0.2cm}&& \hspace{-0.2cm}-
\frac{e^{-(\omega_{K_V} + \omega_{U_V}) t}}{\omega_{K_V}\omega_{U_V}}
-\frac{e^{-(\omega_{K_V} + \omega_{S_V}) t}}{\omega_{K_V}\omega_{S_V}}
\bigg\}, \cr
f_{B}^\kappa(t) \hspace{-0.2cm} &\equiv& \hspace{-0.2cm}{\sum_{\bf k}} \bigg\{
\frac{e^{-(\omega_{K_I} \hspace{-0.05cm}+\hspace{-0.05cm} \omega_{\eta_I}) t}}{6\omega_{K_I}\omega_{\eta_I}}
\hspace{-0.1cm}-\hspace{-0.1cm}\frac{e^{-(\omega_{K_I}\hspace{-0.05cm} +\hspace{-0.05cm} \omega_{U_I}) t}}{2\omega_{K_I}\omega_{U_I}}
\hspace{-0.1cm}-\hspace{-0.1cm}\frac{e^{-(\omega_{K_I}\hspace{-0.05cm} +\hspace{-0.05cm} \omega_{S_I}) t}}{\omega_{K_I}\omega_{S_I}} \cr
\hspace{-0.2cm}&&\hspace{-0.2cm}
+\frac{1}{16}\sum_{b=1}^{16}\left(
2\frac{e^{-(\omega_{K_b} + \omega_{U_b}) t}}{\omega_{K_b}\omega_{U_b}}
\hspace{-0.1cm}+\hspace{-0.1cm}
\frac{e^{-(\omega_{K_b} + \omega_{S_b}) t}}{\omega_{K_b}\omega_{S_b}}\right)
\bigg\}, \nonumber
\label{btt:kappa}
\end{eqnarray}
where $g_{V_{\eta }}$, $g_{V_{\eta'}}$
are denoted in~\cite{Fu:2011xb,Fu:2013sua},
for $f_{A}^\kappa(t)$ we just require $V \to A$ in $f_{V}^\kappa(t)$.
The $\sigma$ bubble term is ~\cite{Fu:2011gw,Fu:2011zz,Bernard:2007qf}
\begin{eqnarray}
B^{\rm S\chi PT}_{2+1,\sigma}(t) \hspace{-0.15cm} &=&
\hspace{-0.15cm} \mu^2/(4L^3) \times \left\{   f_{B}^\sigma(t)
        + f_{V}^\sigma(t)
        + f_{A}^\sigma(t)  \right\} \cr
f_{V}^\sigma(t) \hspace{-0.15cm}&\equiv&\hspace{-0.15cm}
\sum_{\bf k} \bigg\{
-4\frac{e^{-2\omega_{U_V}  t}}{\omega_{U_V}^2}
+C_{V_{\eta }}^2 \frac{e^{-2\omega_{\eta  V} t}}{\omega_{\eta  V}^2}  \cr
\hspace{-0.15cm}&&\hspace{-0.15cm}
+C_{V_{\eta'}}^2 \frac{e^{-2\omega_{\eta' V} t}}{\omega_{\eta' V}^2} -2C_{V_{\eta}}C_{V_{\eta'}}
\frac{e^{-(\omega_{\eta V} + \omega_{\eta' V }) t}}{\omega_{\eta V}\omega_{\eta' V}}
\bigg\}, \cr
f_{B}^\sigma(t) \hspace{-0.15cm}&\equiv&\hspace{-0.15cm} {\sum_{\bf k}} \bigg\{
\frac{1}{9} \frac{e^{-2\omega_{eta I} t}}{\omega_{\eta I}^2}
-\frac{e^{-2\omega_{U_I} t}}{\omega_{U_I}^2} \cr
\hspace{-0.15cm}&&\hspace{-0.15cm} +\frac{1}{16}\sum_{b=1}^{16}
\left(4\frac{e^{-2\omega_{U_b} t}}{\omega_{U_b}^2}
+\frac{e^{-2\omega_{K_b} t}}{\omega_{K_b}^2}\right)
\bigg\}, \nonumber
\label{btt:a0}
\end{eqnarray}
where $C_{V_{\eta }}$, $C_{V_{\eta'}}$
are given
in~\cite{Fu:2011gw,Fu:2011zz,Bernard:2007qf},
for $f_{A}^\sigma(t)$ we just require $V \to A$ in $f_{V}^\sigma(t)$.
In the continuum limit, the surviving thresholds at large $t$ are
\begin{eqnarray}
\hspace{-0.8cm}B^{\rm S\chi PT}_{2+1,\kappa}(t) \hspace{-0.10cm}&=&\hspace{-0.10cm}
\frac{\mu^2}{4L^3} \left\{
\frac{3}{2}\frac{e^{-\left(m_{\pi} + m_{K} \right)t} }{ m_{\pi} m_{K} }
\hspace{-0.10cm}+\hspace{-0.10cm}
\frac{1}{6} \frac{e^{-\left(m_{K} + m_{\eta} \right)t} }{ m_{K} m_{\eta} }
\label{BSchpt_limit_kappa}
\right\}. \\
\hspace{-0.8cm}B^{\rm S\chi PT}_{2+1,\sigma}(t)\hspace{-0.10cm} &=&\hspace{-0.10cm}
\frac{\mu^2}{4L^3} \left\{
3\frac{e^{-2m_\pi t}}{ m_\pi^2} \hspace{-0.10cm}+\hspace{-0.10cm} \frac{e^{-2m_K t}}{ m_K^2}
\hspace{-0.10cm}+\hspace{-0.10cm}
\frac{1}{9} \frac{e^{-2m_\eta t} }{m_\eta^2}
\label{BSchpt_limit_f0}
\right\}.
\end{eqnarray}


\end{document}